\documentclass[conference]{IEEEtran}

\IEEEoverridecommandlockouts
\def\BibTeX{{\rm B\kern-.05em{\sc i\kern-.025em b}\kern-.08em
    T\kern-.1667em\lower.7ex\hbox{E}\kern-.125emX}}

\usepackage[table,xcdraw]{xcolor}
\usepackage{xurl}
\usepackage{subcaption}
\usepackage{svg}
\usepackage{graphicx,pifont}
\usepackage[ruled,linesnumbered,boxed,vlined]{algorithm2e}
\usepackage{algorithmic}
\usepackage{amsfonts}
\usepackage{amsmath,mathtools}
\usepackage{enumitem}
\usepackage{array}
\usepackage{booktabs}
\usepackage{multirow}
\newcommand{\scorebar}[1]{%
  $\ \ \rhd\ \ ${\color[HTML]{41BD35}\rule{
  \dimexpr0.08\dimexpr#1cm-\dimexpr82cm\relax
  }{5pt}}%
}

\SetCommentSty{cmtfont}

\newtheorem{assumption}{Assumption}
\SetKwInOut{Input}{Input}
\SetKwInOut{Output}{Output}

\begin{document}
\title{Ormer: A Manipulation-resistant and Gas-efficient Blockchain Pricing Oracle for DeFi}

\author{\IEEEauthorblockN{Dongbin Bai$^{\textrm{†}}$, Jiannong Cao$^{\textrm{†}}$,~\IEEEmembership{Fellow,~IEEE}, Yinfeng Cao$^{\textrm{†*}}$\thanks{*Yinfeng Cao is the corresponding author.}, Long Wen$^{\textrm{†‡**}}$\thanks{**This work was fulfilled when Long Wen was with The Hong Kong Polytechnic University.}, Milos Stojmenovic$^{\textrm{§}}$}
\IEEEauthorblockA{
$^{\textrm{†}}$The Hong Kong Polytechnic University, Hong Kong, China\\
$^\textrm{‡}$Derivation Technology Limited, Hong Kong, China\\
$^{\textrm{§}}$Singidunum University, Belgrade, Serbia\\
dong-bin.bai@connect.polyu.hk, \{csjcao, csyfcao\}@comp.polyu.edu.hk,\\long.wen@derivation.info, mstojmenovic@singidunum.ac.rs}
}

\maketitle
\thispagestyle{plain}
\pagestyle{plain}

\begin{abstract}
Price feeds of cryptocurrencies are essential for Decentralized Finance (DeFi) applications to realize fundamental trading and exchanging functionalities, which are retrieved from external price data sources such as exchanges and input to on-chain smart contracts in real-time. Currently, arithmetic mean based time-weighted average price (TWAP) oracles are widely used to process price feeds by averaging asset price with short time frame to achieve reliable and gas-efficient pricing. However, recent research indicates that TWAP is vulnerable to price manipulation attacks, resulting in abnormal price fluctuations and severe financial loss. Even worse, TWAP oracles usually set a relatively long time frame setting to prevent such attack. However, it would further introduce long delays and high price deviation errors from the market asset price.

To address this issue, we propose a novel on-chain gas-efficient pricing algorithm (\textsc{Ormer}) that heuristically estimates the median of asset price within an observation window based on a piecewise-parabolic formula, while the time delay is suppressed by fusing estimations with different window sizes. Our evaluation based on multiple pairs of token swapping price feed across different chains show that \textsc{Ormer} reduces the mean absolute price error by $15.3\%$ and the time delay by $49.3\%$ compared to TWAP. For gas efficiency, regardless of the number of price observations, an encoding mechanism with constant storage requirement is employed without saving all the historical data for median estimation. Surprisingly, the lowest gas consumption of \textsc{Ormer} is even $15.2\%$ less than TWAP, and the oracle querying fee would be saved up to $\sim$20K USD per day for DeFi participants.
\end{abstract}


\section{Introduction}

\begin{figure}[ht!]
\vspace{1em}
\centering
\includegraphics[width=0.38\textwidth]{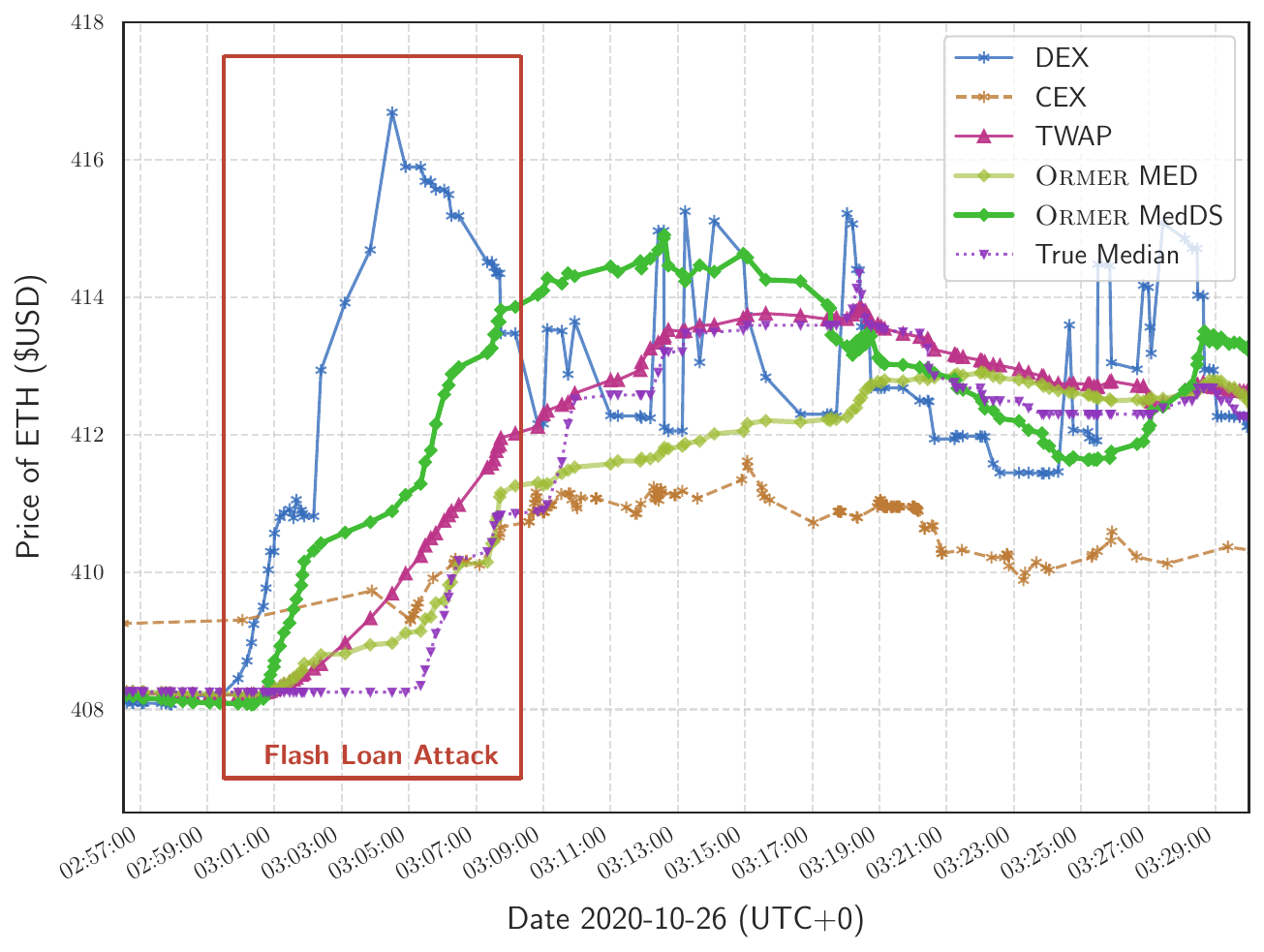}
\caption{A real-world flash loan-based price manipulation attack \cite{eminence2020attack} that makes DEX price feed fluctuates abnormally. TWAP price feed achieves fair manipulation resistance to price fluctuation but fails to reflect the latest market price. \textsc{Ormer} reflects the market price better by following closely to the trend of DEX and CEX price feeds at the same time, while introducing fair fluctuation and quick recovery speed after the attack. }
\label{fig:ormer}
\vspace{-1.3em}
\end{figure}

In recent years, blockchain-based Decentralized Finance (DeFi) applications have made significant advances in reshaping traditional financial services. Empowered by on-chain smart contracts, these applications have provided a fair and open financial platform for different stakeholders \cite{werner2022sok}. For example, various novel financial applications and services are proposed, such as token swapping (exchange) \cite{uniswapv3,heimbach2022exploring}, lending \cite{Aave,Compound}, and derivatives\cite{Synfutures}, allowing users and institutions to conduct diverse financial businesses flexibly with tokens. By 2025, there are over $\$108$ billion Total Value Locked (TVL) in over $3,800$ DeFi applications on various blockchains like Ethereum, Binance Smart Chain, and Solana, indicating that a new global decentralized finance ecosystem is rapidly forming \cite{DefiLlama}.

Among which, pricing data is essential for DeFi applications. It reflects continuous and accurate cryptocurrency/fiat money market information as important reference for on-chain applications like trading, swapping, loaning, and depositing, etc. For example, the exchange rate between ETH and USD is frequently utilized by on-chain token swapping applications for reference to trade various tokens on Ethereum.

\textbf{Price Feed Processing.} Due to the decentralized nature of blockchains and smart contracts (automated on-chain program), processing such pricing data needs multiple complex procedures. Specifically, in practice, such price data are originally generated from \textit{Price Data Sources}, such as off-chain Centralized Exchanges (CEXs), Decentralized Exchanges (DEXs), and Automated Market Makers (AMMs), where trading and arbitraging activities happen and decide the market price \cite{fritsch2021concentrated,ni2024money}. To this end, \textit{blockchain pricing oracle} is developed as a crucial on-chain contract to process these original market price data into a correct and continuous form (\textit{feed}), which is further fed to downstream DeFi applications.

At current stage, \textit{time-weighted average price} (TWAP)-based pricing oracle is the most popular pricing oracle developed by Uniswap \cite{uniswapv3}, which simply takes arithmetic mean of the token quantities collected from DEXs with a pre-set fixed time frame for updating price data. As only a minimal storage space in smart contract is required, TWAP demonstrates a high gas-efficiency in practical deployment. Compared with centralized solutions like Chainlink, TWAP oracle is fully on-chain and thus considered more trustworthy. As a result, TWAP oracles have been adopted as the mainstream price references (around $60$\% market share for DEX token pricing) for DeFi applications.

\textbf{Problem.} However, many recent studies have found that TWAP oracles severely suffer from manipulation attacks such as flash loan attacks and arbitrage attacks \cite{mackinga2022twap,koutsoupias2019blockchain}. Specifically, for a short time frame setting of a TWAP oracle, attackers could make the oracle output price data deviates by injecting extreme price data points through multiple malicious transactions (e.g., use an extremely low or high rate to swap tokens). Even worse, for a TWAP oracle with a long time frame for mitigation of manipulations, the time delay and price deviation error to market price would be increased because the oracle fails to follow the up-to-date price, while a higher price deviation error would create additional opportunities for arbitrage attacks by malicious parties \cite{heimbach2024non,wang2022cyclic}. According to the top $200$ costliest attacks recorded in Rekt Database, the financial losses caused by only $36$ flash loan attacks have exceeded $418$ million USD \cite{mckay2022defi, zhou2023sok,kulkarni2023routing}.

\textbf{Challenges.} Different from traditional financial services, on-chain pricing data can not be directly used by DeFi applications because they are (1) \textit{Discrete:} price data points are usually generated and changed block-by-block (e.g., the price from on-chain DEX or AMM); (2) \textit{Deviated:} price data points are only updated by public invocations from users or contracts; (3) \textit{Expensive:} smart contract invocations are costly in public blockchain, making DeFi participants less willing to update the pricing data. These factors make data points scattered or out-dated, and can hardly reflect the market price correctly in time (\S\ref{sec_challenges}).

In this paper, we propose \textsc{Ormer}, the first manipulation-resistant and gas-efficient on-chain pricing oracle for DeFi. In particular, inspired by traditional financial stock pricing services, \textsc{Ormer} novelly adopts the median-based method to process on-chain price data from a market source, which is effective on resisting manipulations as the price median is hard to be affected by few extreme prices. Moreover, we overcome the expensive gas consumption issues of calculating streaming median in smart contracts by a carefully designed algorithm, which only requires a constant storage space by heuristically constructing the median estimations of a past sliding window. We also conduct extensive experiments and practical deployment to demonstrate the improved security, performance, and effectiveness in practice. To emphasize, the algorithm is a general alternative to TWAP, which can also be used in traditional financial data management services.

\noindent\textbf{Contributions}. The contributions of \textsc{Ormer} can be summarized as follows:

\begin{itemize}[left=0pt]
    \item \textbf{\textsc{Ormer}:} in this paper, we propose the first on-chain pricing algorithm for blockchain pricing oracle with SotA performance, providing the following properties:
    \begin{itemize}
    \item \textbf{Streaming Median:} \textsc{Ormer} could estimate the median of the last sliding window of streaming price feed in a gas-efficient manner without storing all the historical data in a fully on-chain setting (\S \ref{subsec_medianestimator}).
    
    \item \textbf{Manipulation Resistant:} \textsc{Ormer} estimates the median of a given input price feed window, which significantly raises the cost on price manipulation attack, since more corrupted prices are needed to affect the feed (\S \ref{subsec_medianestimator}). According to the metrics defined in \S\ref{subsec_metrics}, decentralized \textsc{Ormer} MedDS (one derivative of \textsc{Ormer} requires more gas for a low delay) achieves $90.0\%$ performance of True Median and $65.7\%$ performance of a centralized oracle regarding manipulation resistance, with $75.4\%$ improvement to TWAP.

    \item \textbf{Delay Reduction:} with the method introduced in \S \ref{subsec_delay}, price feed constructed by \textsc{Ormer} MedDS could achieve significant lower time delay than other smoothing based methods. The delay is reduced by $49.3\%$ compared to TWAP. Statistical results indicate that purely on-chain \textsc{Ormer} MedDS could reach only $4\%$ more delay in time in fully decentralized manner compared to centralized off-chain oracle (\S\ref{sec_eval}).

    \item \textbf{Gas Efficient:} \textsc{Ormer} MED (another derivative of \textsc{Ormer} requires less gas for accurate estimation of median) requires a fixed $256$ bits of contract storage, while $512$ bits of fixed storage is required for \textsc{Ormer} MedDS with delay suppression (\S \ref{subsec_slot}). Both modes consume less gas on oracle price querying compared to TWAP, which would ultimately save up to $\$19$ USD per querying. In real-world environment, \textsc{Ormer} MED could help DeFi participants save up to $\$26$k USD per day (\S \ref{sec_eval}).
    \end{itemize}
    \item \textbf{Implementation and Compatibility:} we implement \textsc{Ormer} as an example oracle contract in Solidity language, which is deployed in real-world environment\footnote{\url{https://sepolia.etherscan.io/address/0x02fe8289781ec42ab64408720328667cca820026}}. \textsc{Ormer} is compatible to scale on most of the mainstream blockchains (e.g., Ethereum Virtual Machine compatible blockchains like Ethereum, Tron, Polygon) without needing significant modifications. All the source codes and experimental artifacts are made publicly available\footnote{\url{https://github.com/EvanBin/Ormer}}.

    \item \textbf{Experimental Evaluation:} we have collected $4,130,000$ spot prices from real-world swapping pools (DEXs) from 2022-01-13 to 2023-09-13 as workloads. To the best of our knowledge, we are the first to conduct quantitative experiment on blockchain oracles. We comprehensively compare \textsc{Ormer} MED and \textsc{Ormer} MedDS with Chainlink, TWAP, Exponential Moving Average (EMA), and True Median with practical implementations. We also firstly define several metrics for evaluating the pricing oracles. Results indicate that both modes of \textsc{Ormer} outperform existing widely deployed on-chain pricing oracles (\S \ref{sec_eval}).
\end{itemize}

\section{System Model}\label{sec_problem}

In most DeFi applications requiring price feed for trading, exchanging, etc, there are three main components over time $t$ as shown in Figure \ref{fig:flash_loan}: 

\begin{itemize}[left=0pt]
    \item A price data source (DEX, CEX, AMM, etc.) $SC$ that actually produces a source price feed $\mathcal{P}_{sc}=\{(t_1,p_1^{sc}),...,(t_n,p_n^{sc})\}$ of spot price (i.e., the latest price for successful orders) for the rate between Token $B$ and Token $A$ (e.g, USDT and ETH). At $t$, the market price is $\mathcal{P}_{market}=\{(t_1,p_1^{market}),...,(t_n,p_n^{market})\}$.
    \item A pricing oracle $PO$ that takes one $\mathcal{P}_{sc}$ as input and output oracle price feed $\mathcal{P}_o=\{(t_1,p_1^o),...,(t_n,p_n^o)\}$.
    \item A DeFi application $APP$ that receives $\mathcal{P}_o$ as reference. 
\end{itemize}

\subsection{Price Data Source}

A price data source $SC$ can be an AMM, DEX, or CEX, which provides a source price feed $\mathcal{P}_{sc}=\{(t_1,p_1^{sc}),...,(t_n,p_n^{sc})\}$ according to the successful orders. Therefore, price feeds from all price data sources in market should reflect a logical market price feed $\mathcal{P}_{market}=\{(t_1,p_1^{market}),...,(t_n,p_n^{market})\}$. 

For example, AMM is typically implemented by a smart contract that provides token swapping service between Token $A$ and Token $B$ (e.g., USDT and ETH) according to the liquidity pool, and the price feed $\mathcal{P}_{sc}$ from AMM can be accessed directly by DeFi applications.
Similarly, DEXs can be composed by AMMs, and a CEX's price data can be relayed to blockchains block-by-block, they all have identical features with general DEXs under this system model. 

However, due to the limited liquidity of any single price data source, the price feed $\mathcal{P}_{sc}$ may deviate from the logical market price feed $\mathcal{P}_{market}$, especially when manipulation attacks happen.

\subsection{Pricing Oracle}
Therefore, to provide a more stable and continuous \textit{price feed} based on price data feed from price data source, Pricing Oracle is proposed and deployed on-chain between the price data source $SC$ and the DeFi application $APP$.

A pricing oracle $PO$ takes source price feed $\mathcal{P}_{sc}=\{(t_1,p_1^{sc}),...,(t_n,p_n^{sc})\}$ to construct a new price feed $\mathcal{P}_o=\{(t_1,p_1^o),...,(t_n,p_n^o)\}$ by running a pricing algorithm.

Ideally, $\mathcal{P}_{sc}$ should have no deviation from the market price feed $\mathcal{P}_{market}$ if the smart contract smoothly updates in real-time. However, as the $\mathcal{P}_{sc}$ is produced discretly block-by-block (several seconds on blockchains) instead of real-time (milliseconds in traditional exchanges), it may be significantly deviated from $\mathcal{P}_{market}$ across different blocks. Moreover, as $PO$ is typically implemented by a smart contract, it is extremely costly to update price feed in real-time.  Therefore, the design of pricing algorithm inside $PO$ is the key, which should be designed to be robust (even against the manipulation attacks) and gas-efficient.

\subsection{DeFi Application}
Therefore, for safety considerations, DeFi applications typically read price feed $\mathcal{P}_{o}$ from $PO$ instead of directly reading $\mathcal{P}_{sc}$ from $SC$ to get the latest and robust market price.

For example, loaning application is a mainstream type of DeFi application that accepts cryptocurrency as collateralization. It is important for such application to get the latest market price of the collateralized cryptocurrency for its own profit (i.e., liquidize the collateralized cryptocurrency when its value drops). As shown in Figure \ref{fig:flash_loan}, if the pricing oracle feed is influenced by the manipulated Price Data Source, the liquidation process designed to protect the application profit will be exploited against the application owner by false triggering or not triggering \cite{zhou2023sok}. Similarly, such pricing algorithm is also essential and desired by other DeFi applications including token swapping, derivatives trading, yield farming (deposit cryptocurrency to a platform with investing strategy for better revenue), etc \cite{werner2022sok}.

\subsection{Threat Model}

\noindent \textbf{Price Manipulation Attack.} In practice, price data sources with low liquidation are vulnerable to price manipulation attacks \cite{mckay2022defi}, where attacker aims to affect a DEX's spot price by artificially determine the spot price of a token with one large volume transaction. As results, these attacks would lead to severe financial losses in real-world scenarios \cite{mackinga2022twap,xi2024pomabuster,qin2021attacking}.

\begin{figure}[!t]
    \centering
    \includegraphics[width=0.98\linewidth]{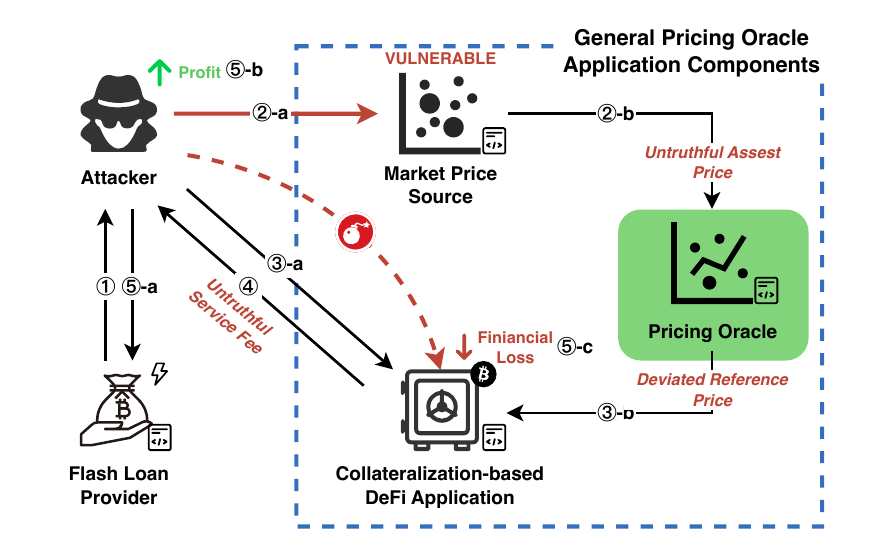}
    \caption{A typical flash loan based price manipulation attack. An attacker manipulates the \textit{Price Data Source} with flash loan (\textcircled{1}\textcircled{2}); A downstream application (i.e., \textit{Collateralization-based DeFi application}) takes manipulated Oracle Price Feed (constructed by a \textit{Pricing Oracle}) to determine financial service fee (\textcircled{3}\textcircled{4}); The attacker profits from untruthful service fee: here, undercollateralized loan (\textcircled{5}).}
    \label{fig:flash_loan}
\end{figure}

\noindent\textbf{Example.} As shown in Figure \ref{fig:flash_loan}, a typical construction of flash loan-based price manipulation attack aims to cheat the downstream DeFi applications with false pricing data to make profit. Here we provide an example of flash loan based price manipulation attack, which involves manipulating the price between two DeFi applications:

\begin{itemize}[left=0pt]
    \item\textit{Flash Loan Provider}: smart contract $APP_{fp}$ that offers token liquidity without any prior collateral requirement.

    \item\textit{Collateralization-based DeFi application}: smart contract $APP_{Col}$ that offers Token $A$ loaning service by collateralizing Token $B$.
\end{itemize}

The attacker first borrows $10,000$ Token $A$ from $APP_{fp}$ with no prior collateralized assets, which is made possible based on the atomicity property of blockchain transactions (\textcircled{1}). Then the attacker would use the borrowed Token $A$ to swap for $1,000$ Token $B$ in $APP_{fp}$, which would significantly raises Token $B$'s spot price $10\ A/B$ into $100\ A/B$, since the saved quantity of Token $A$ in the exchange is pushed up to an abnormally high position (\textcircled{2}-a). Under this situation, if $PO$ fails to smooth the abnormal surge of spot price, the oracle referencing price $9.8\ A/B$ would deviates into $97\ A/B$ (\textcircled{2}-b). At this moment, the attacker invokes service of borrowing Token $A$ by collateralizing Token $B$ in a downstream application $APP_\text{Col}$ (\textcircled{3}-a). To determine the price of Token $B$, the application $APP_\text{Col}$ would query $PO$ for latest oracle referencing price, resulting in being cheated by the attacker that the price of Token $B$ has dramatically increased (\textcircled{3}-b). Thus, $APP_\text{Col}$ lends $12,000$ Token $A$ to the attacker, as it is cheated to have an over-collateralization of Token $B$ (\textcircled{4}). In the end, the attacker pays back borrowed $10,000$ Token $A$ to $APP_{fp}$ and default the loan from $APP_\text{Col}$ (\textcircled{5}-a). As a result, the attacker earns $2,000$ Token $A$ (\textcircled{5}-b). 

In this example, the Collateralization-based DeFi application $APP_\text{Col}$ suffers from an under-collateralization situation with financial loss, which is caused by a pricing oracle that fails to smooth the abnormal surge of spot price (\textcircled{5}-c) \cite{mckay2022defi}.

\begin{figure*}[!t]
\centering
\includegraphics[width=0.85\textwidth]{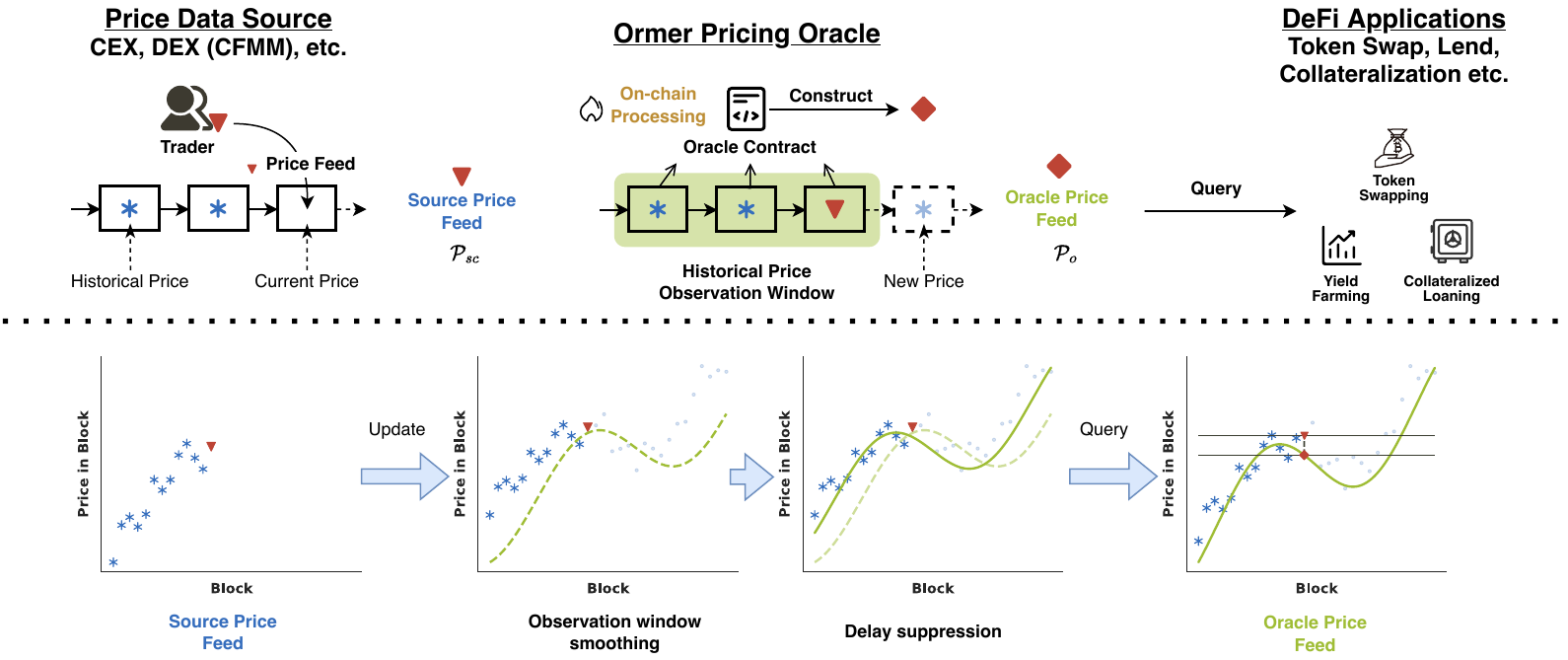}
\caption{Overview of \textsc{Ormer} system model workflow with three main components: Market Price Source, Pricing Oracle, and DeFi Protocol. Instead of relying potentially vulnerable Spot Price Feed, DeFi Protocol would use Pricing Oracle Feed that is constructed upon weighted historical data with delay suppression techniques.}
\label{fig:system}
\end{figure*}

\section{Motivations and Objectives}\label{sec_challenges}

To counter price manipulation and abnormal price fluctuations, TWAP oracles are widely adopted in practice, which is designed to smooth the price feed by averaging the incoming price data over a fixed time window. However, many recent studies have found that TWAP oracles actually failed \cite{mackinga2022twap,koutsoupias2019blockchain}. Specially, as illustrated in Figure \ref{fig:flash_loan}, for a short time frame setting of TWAP oracle, it is ineffective to filter out the extreme price data points injected by attackers. For a TWAP oracle with a long time frame, the time delay and price deviation error would be also increased because the oracle fails to follow the up-to-date delay, while a higher price deviation error would create additional opportunities for arbitrage attacks by malicious parties \cite{heimbach2024non,wang2022cyclic}.

Other price feed smoothing methods, such as median-based algorithms that are widely used in traditional trading systems, are proved to be effective for smoothing the price feed and filtering out the extreme price data points. However, they require a large amount of storage space to store the historical price data, which is impractical for on-chain implementation due to the high gas consumption for storage. For example, querying a price data point from a median-based oracle would require around three times more gas ($\sim$310k gas) than a TWAP oracle ($\sim$100k gas).

Therefore, we have following objectives when designing \textsc{Ormer}:
\begin{itemize}[left=0pt]
    \item \textbf{Robust Deviation Resistance}: The pricing oracle should be robust against manipulation attacks, which means the deviation between the pricing oracle feed and the market price feed should be minimized.
    \item \textbf{Low Delay}: The pricing oracle should be able to provide the latest market price in real-time.
    \item \textbf{Low On-chain Cost}: The pricing oracle should be gas-efficient, which means the gas consumption of the pricing oracle should be minimized.
\end{itemize}

\subsection{Robust Deviation Resistance}
The primary objective of pricing oracle is to provide a price feed $\mathcal{P}_{o}$ according to delayed $\mathcal{P}_{sc}$ that has minimal deviation from $\mathcal{P}_{market}$ in a fully on-chain manner. Formally, given the time $t\in\mathcal{T}$, true external price feed $\mathcal{P}_{market}=\{(t,p_t^{market})|t\in\mathcal{T}\}$, exchange price feed $\mathcal{P}_{sc}=\{(t,p_t^{sc})|t\in\mathcal{T}\}$, the deviation with observation window $L$ can be defined as:

\begin{equation}\label{eq:oracle_constrain_distance}
\resizebox{0.98\linewidth}{!}{$
    \begin{aligned}
    \mathcal{D}(\mathcal{P}_{sc},L):=\mathop{min}\limits_{\mathop{t\in\mathcal{T},}\limits_{n_0=|\mathcal{T}|-L}} \sum_{n_0}^{\mathcal{N}}\sum_{t_0}^{\mathcal{T}_L}Distance(Oracle(p_t^{sc})^2-{p_t^{market}}^2)
    \end{aligned}
$}
\end{equation}

In this paper, we assume that only upper bound of the manipulated prices is known. Since the oracle contract do not have direct access to $\mathcal{P}_{market}$, it could be impossible for the contract to construct $\mathcal{P}_{o}$ when all the data of $\mathcal{P}_{sc}$ are manipulated, so an upper bound $\beta$ is assumed. 

\begin{assumption}
An adversary $e_t$ at time $t$ arbitrarily manipulates at most $\beta$ number of $\ p_{t'}$ in observation window with size $L$ of $\mathcal{P}_e$.
\end{assumption}

Given a back-testing evaluation function $\mathcal{V}$, we say $\mathcal{V}$ is $(\beta,\epsilon)$-secure if it satisfies:

\begin{equation}
\mathcal{V}(\mathcal{P}_o, \mathcal{P}_{market}, \beta, L)\leq \epsilon
\end{equation}

\subsection{Low Delay} While robust estimations can effectively resist manipulation attack, they may fail to accurately reflect the latest market dynamics of the external centralized exchanges and the market price. The underlying principle behind a pricing oracle is to sample a subset of historical data for generating a statistical result as the current price. This approach inherently hinders the timeliness of the output. Specifically, a naive way to convert the discrete DEX price feed into a continuous and more dense and stable feed is by employing a moving average filter, which is the underlying principle used by TWAP. Oracle price calculated according to such filter would falls behind DEX in time, varies by the choice of averaging window size. In order to keep pace with delay, we empower \textsc{Ormer} with prediction functionality that fuses two parallel updated estimations as the final delay suppressed output.

\subsection{Low On-chain Cost}

The primary challenge \textsc{Ormer} faces is to design a purely on-chain pricing oracle. Although there are many robust streaming data estimation methods studied in data science (e.g., t-digest, Quantile Regression, Gaussian Processes, etc.), the implementation of such methods are gas-consuming in smart contract due to their complex matrix computations, sophisticated data structures, or large storage space requirements for historical data. Since the pricing oracle is updated upon contract invocations, directly migrating such methods to blockchain is impractical. For example, about $\$6$ USD for accessing a $256$ bits value in Ethereum (estimated according to: $3,000$ USD/ETH, $100$ Gas Price). Thus, it is desired to have a pricing oracle with less smart contract storage requirement and computation steps. Formally, given the time $t\in\mathcal{T}$, true external price feed $\mathcal{P}_{market}=\{(t,p_t^{market})|t\in\mathcal{T}\}$, exchange price feed $\mathcal{P}_{sc}=\{(t,p_t^{sc})|t\in\mathcal{T}\}$, the gas consumption with observation window $L$ can be defined as:

\begin{equation}\label{eq:oracle_constrain_gas}
\mathcal{G}(\mathcal{P}_{sc},L):=\mathop{min}\limits_{n_0=|\mathcal{T}|-L} \sum_{n_0}^{\mathcal{N}}Gas(Oracle(\mathcal{P}_{sc}|t\in[n,n+L-1]))
\end{equation}

\subsection{Decentralization}

Centralized solutions such as Chainlink can provide a low-deviation price feed with low latency and at a low cost by using high-end servers to periodically fetch data from several of the largest CEXs. However, for DeFi applications built on such centralized solutions, a connection failure with the servers would result in business failure, thereby exposing severe attack surfaces to malicious parties. Furthermore, it is hard to ensure or verify whether the servers have been corrupted by external attackers or internal operators. Therefore, they are not utilized by the most of DeFi applications in practice \cite{ni2024money, zhou2023sok, lee2020measurements}.

\section{\textsc{Ormer} Pricing Oracle}\label{sec_ormer}

\subsection{Overview}

In order to provide $\mathcal{P}_o$ with security insurance while remaining gas consumption comparable to State-of-the-Art TWAP (one cold slot read per update), we propose a novel streaming median estimation algorithm: \textsc{Ormer}. The estimation feed $\mathcal{P}_o$ is produced dynamically, which could follow the trend of input source with less time delay as the spot price $\mathcal{P}_e$ changes, and the algorithm has a constant storage requirement regardless of the number of price observations. The goals are achieved by three main components:

\begin{itemize}[left=0pt]
    \item \textbf{Median Estimator} (\S\ref{subsec_medianestimator}): we design a sliding window median estimation algorithm that accepts data feed input in a streaming manner. The algorithm does not require to store all the historical data for median output by using a piecewise-parabolic formula and an auxiliary sliding storage for median estimation.
    \item \textbf{Delay Suppression} (\S\ref{subsec_delay}): we propose a novel price feed delay suppression method to better reflect market price is achieved by utilizing the projection angle of the \textit{markers}.
    \item \textbf{Slot Encoding} (\S\ref{subsec_slot}): we design a novel slot encoding template for smart contract implementation that would reduce the gas consumption.
\end{itemize}

\subsection{Median Estimator}\label{subsec_medianestimator}
Inspired by \cite{akinshin2024trimmed,jain1985p2}, we have developed a novel sliding window estimator that could deal with the streaming input data with only five \textit{markers}: the minimum, the 25th percentile, the 50th percentile, the 75th percentile and the maximum price in the observation window. A \textit{marker} is a vertical stick that contains $(n_i,h_i)$, where $n_i$ is the horizontal position of stick and $h_i$ is the vertical height of stick. With five sticks sorted from the lowest height to the highest height, stick with index $2$ (start from $0$) is regarded as the estimation of median based on the recorded states. When a new $p_t^e$ observation comes in, it is compared with stored markers. As a result, all the markers with greater heights than observation are moved one position to the right (denoted as $n_i+1$). After $n$ observations, the desired position $n'_i$ of markers should be: $n'_0=0,\ n'_1=\frac{n}{4},\ n'_2=\frac{n}{2},\ n'_3=\frac{3n}{4},\text{ and }n'_4=n$.

\begin{figure}[t!]
\centering
\includegraphics[width=0.8\columnwidth]{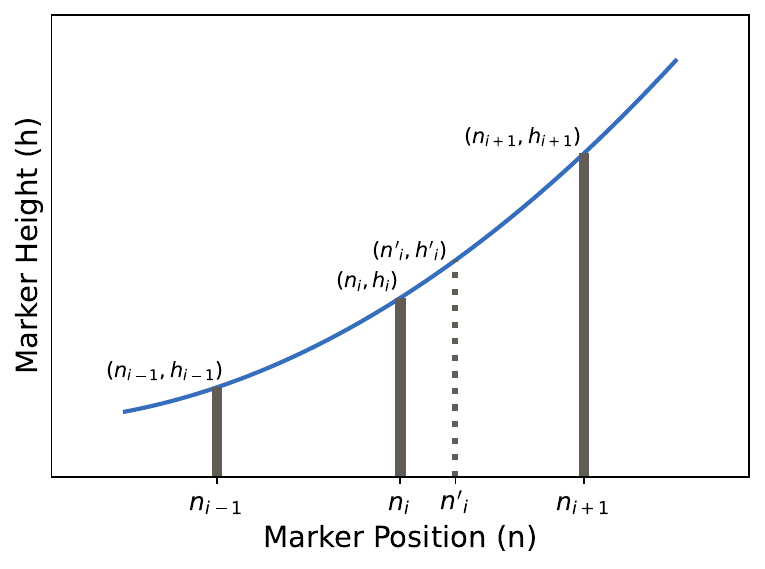}
\caption{Marker update process of $n'_i$ with parabolic formula.}
\label{fig:marker}
\end{figure}

If the moved marker position $n_i$ deviates its ideal position $n'_i$ more than one, it means the distribution of observation may have shifted from the states stored in smart contract. To follow up the current distribution, both the position and height should be adjusted. For the position, we would prefer to move it either one to the right ($+1$) or one to the left ($-1$) determined by the sign of $d_i=n'_i-n_i$. For the height adjustment, as shown in Figure\ref{fig:marker}, markers with position $[n_0,n_1,n_2]$, $[n_1,n_2,n_3]$, and $[n_2,n_3,n_4]$ would be updated sequentially when a new observation arrives according to parabola with form $h_i=an_i^2+bn_i+c$. It is necessary to mention that $a$, $b$, and $c$ are coefficients determined by three position-height pairs: $(n_{i-1},h_{i-1})$, $(n_{i},h_{i})$, and $(n_{i+1},h_{i+1})$, where $i\in\{1,2,3\}$. According to the position index $i-1$, $i$, and $i+1$, given ${n'}_i=n_i+d_i$, it is straightforward to get ${h'}_i$ with ${h'}_i=a{n'}_i^2+b{n'}_i+c$:

\begin{equation}\label{eq:pp}
\resizebox{0.9\linewidth}{!}{$
    \begin{aligned}
    \begin{split}
    {h'_i} & = a(n_i+d_i)^2 + b(n_i+d_i) + c \\
    & = \!\begin{multlined}[t]
    h_i + \frac{d_i}{n_{i+1}-n_{i-1}} \\
    \cdot \left[ (n_{i}-n_{i-1}+d_i) \frac{h_{i+1}-h_{i}}{n_{i+1}-n_{i}} + (n_{i+1}-n_{i}-d_i) \frac{h_{i}-h_{i-1}}{n_{i}-n_{i-1}} \right]
    \end{multlined}
    \end{split}
    \end{aligned}$}
\end{equation}

\begingroup
\setlength{\textfloatsep}{1em}
\begin{algorithm}[ht!]
    \caption{\textsc{Ormer MED}}\label{alg:ormer}
    \Input{External price feed $\mathcal{P}_e=\{(t_{1},p_{1}^e),...,(t_{n},p_{n}^e)\}$}
    \Output{Oracle price feed $\mathcal{P}_o=\{(t_{1},p_{1}^o),...,(t_{n},p_{n}^o)\}$}

    $\text{count} \gets 0\text{\; last estimation} \gets 0$\;
    $n[5] \gets [1,2,3,4,5]\;\ h[5] \gets [0,0,0,0,0]$\;
    $dn[5]=[0,0.25,0.5,0.75,1.0]$
    
    \While{$\mathcal{P}_e\neq \phi$} {
    $p_t\gets \text{latest}\ p_t^e\ \text{at time}\ t\text{\; count}\gets \text{count} + 1$\;
        \If{$\text{count} == $Observation Window Size} {
            Update last window estimation, initialize markers\;
        }
        
        \If{$\text{count} < 6$} {
            $\text{Latest non-initiated marker height} \gets p_t$\;
            \uIf{$\text{count} == 5$} {
                $\text{Latest non-initiated marker height} \gets p_t$\;
                Sort ascending marker heights $h[5]$\;
            }
            \Else {
                continue\;
            }
        }
        \tcp{Find cell $k$ that $h_k\leq p_j \leq h_{k+1}$ and adjust $h_0$ and $h_4$ if selected}
        \Switch{$p_t$} {
            \lCase{$p_t<h_0$}{$h_0\gets p_t\text{\; }k\gets 0$}
            \lCase{$h_0\leq p_t\leq h_1$}{$k\gets 0$}
            \lCase{$h_1\leq p_t\leq h_2$}{$k\gets 1$}
            \lCase{$h_2\leq p_t\leq h_3$}{$k\gets 2$}
            \lCase{$h_3\leq p_t\leq h_4$}{$k\gets 3$}
            \lCase{$h_4\leq p_t$}{$h_4\gets p_t\text{\; }k\gets 3$}
        }
        $n_i\gets n_i+1\quad i=k,...,4\text{\; }n'_i\gets \text{count} \cdot dn[i]\quad i=0,...,4$\;
        
        \tcp{Marker height adjustment}
        \For{$i = 1;\ i < 4;\ i = i + 1$}{
            $d_i\gets n'_i-n_i$\;
            Determine sign of $d_i$, position of $n_i$ would either move $+1$ or $-1$\;
            Try $h'_i \gets h_i$ from parabolic formula\;
            \uIf{not $h_{i-1} < h'_i < h_{i+1}$} {
                $h_i\gets h'_i$\;
            }
            \Else{
                $h'_i \gets h_i$ from linear formula\;
            }
            $n_i\gets n_i+d_i$\;
        }
        
        $p_t^o\gets$ Estimates according to $h[2]$ and last window estimation\;
    }
\end{algorithm}
\begingroup

For the algorithm to work correctly, the heights of markers should remain in an ascending order ($h_{i+1}\geq h_i,\  i\in\{0,1,2,3\}$). So if an unlawful estimation $h'_i$ that locates outside desired range $(h_{i-1}<h'_i<h_{i+1})$, the result is ignored and $h'_i$ is recalculated according to linear formula where $d_i=\operatorname{sign}(n_i' - n_i)$:

\begin{equation}
h'_i=h_i+d_i\frac{h_{i+d_i}-h_i}{n_{i+d_i}-n_i}, \quad d_i \in \{1,-1\}
\end{equation}

\noindent \textbf{\textsc{Ormer}-Median.} With the marker updating process described above, we are able to construct \textsc{Ormer}-Median that constantly takes input $\mathcal{P}_e$ from time $0$ to $t$ and estimates the median of all the received data as $p^o_{0,t}$. However, \textsc{Ormer}-Median is not capable of estimating the median of given window size in an online streaming data setting (e.g., window time from $t_1$ to $t_2$, expect $p^o_{t_1,t_2}$).

\noindent \textbf{\textsc{Ormer}-SlidingWindow.} To resolve the flaw, we need to add an additional storage requirement for recording the median estimation result of the last window $E_{\text{last}}$. \textsc{Ormer}-SlidingWindow is designed to receive and deal with streaming data using sliding window. First, an \textsc{Ormer}-Median is initialized to receive coming $p_t^e$. Given the manually defined window size $L$, once the number of $L$ is fulfilled, $E_{\text{last}}$ is estimated as $h_2^{\text{last}}$ \textsc{Ormer}-Median according to historical data collected in last window. Then \textsc{Ormer}-Median is reset to initial state. For each subsequent \textsc{Ormer}-SlidingWindow estimation $E_t$, it could be approximated with weighted sum:

\begin{equation}
E_t=\frac{(L-c_t) \cdot E_{\text{last}}+c_t \cdot E_{\text{current}}}{L}
\end{equation}

\noindent where $c_t$ is the number of observation received of current window, and $E_{current}$ is given by $h_2^{\text{current}}$. In the first window, the $E_t$ would be simply given by $h_2$ of \textsc{Ormer}-Median. \textsc{Ormer} MED is the combination of \textsc{Ormer}-Median and \textsc{Ormer}-SlidingWindow, which use $E_t$ as $p_t^o$, and the pseudocode of our proposed algorithm is summarized in Algorithm \ref{alg:ormer}.

\begin{figure}[t!]
\centering
\begin{subfigure}[t]{0.222\textwidth}
\centering
\includegraphics[width=\linewidth]{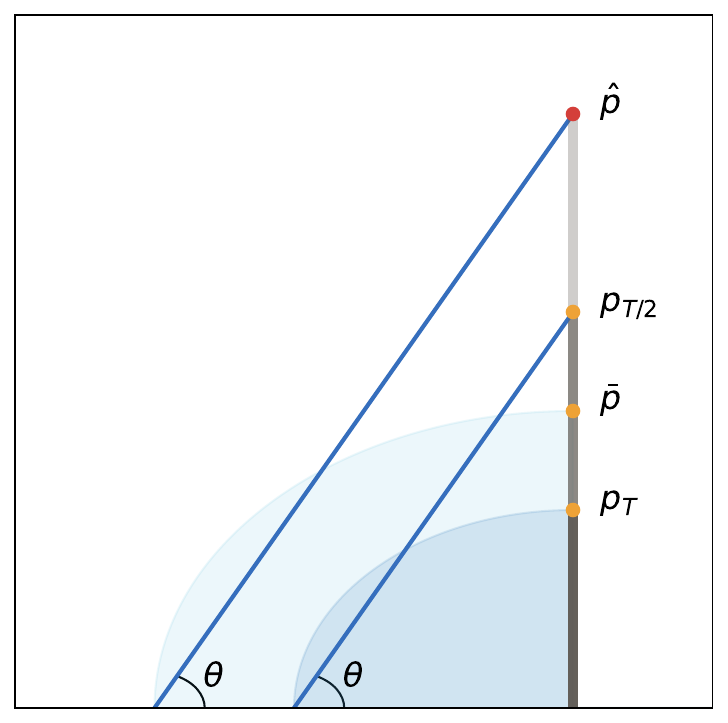}
\caption{Estimate $\hat{p}$ when $p_{T/2}$ marker is longer than $p_{T}$ marker.}
\label{fig:scale1}
\end{subfigure}
\hfill
\begin{subfigure}[t]{0.2222\textwidth}
\centering
\includegraphics[width=\linewidth]{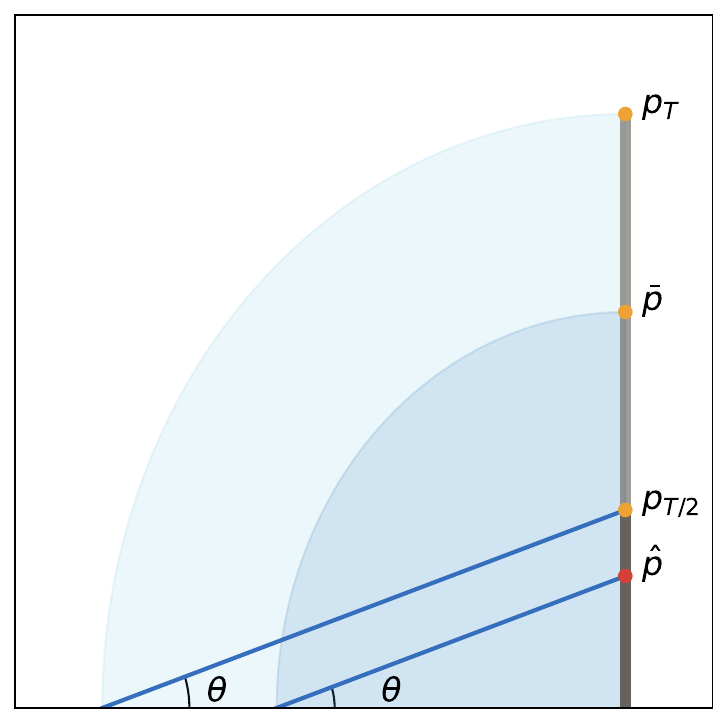}
\caption{Estimate $\hat{p}$ when $p_{T}$ marker is longer than $p_{T/2}$ marker.}
\label{fig:scale2}
\end{subfigure}
\caption{Estimate non-delayed data $\hat{p}$ with $p_{T/2}$ and $p_{T}$.}
\label{fig:scale}
\end{figure}

\subsection{Delay Suppression}\label{subsec_delay}

It is unavoidable estimating current time data point based on the historical observations. In this paper, we propose to fuse additional information from a smaller observation window $L=T/2$ besides updating \textsc{Ormer} MED according to manually settled $L=T$. Intuitively, estimation of $p_{T/2}$ would provide more up-to-date information since it is constructed based on historical data ranging from $t-(T/2)$ to $t$ compared to $t-T$ to $t$. Simply averaging these two information with $\bar{p}=(p_{T/2}+p_{T})/2$ would let estimated $\hat{p}$ always fall between $p_{T/2}$ and $p_{T}$. As shown in Figure \ref{fig:scale}, in order to better estimate non-delayed data $\hat{p}$, we rotate $p_T$ marker to x-axis, forming a new projection line ${\tilde{p_T}p_{T/2}}$ with slope $\tan\theta=p_{T/2}/p_{T}$. For estimation of $\hat{p}$, we assume if $\bar{p}$ marker is rotated to x-axis, it would share the same slope $\theta$. Thus, estimation of $\hat{p}_t$ at time $t$ is given by:

\begin{equation}\label{eq:delaysuppress}
\hat{p_t} = \frac{p_{T/2}+p_{T}}{2} \cdot \frac{p_{T/2}}{p_{T}}
\end{equation}

As demonstrated in Figure \ref{fig:scale1} and Figure \ref{fig:scale2}, $\hat{p}$ would aggressively follow the trend (either increase or decrease) of $p_{T/2}$ while fusing the more smoothed $p_T$ to ensure the property of manipulation resistance. With the design of Eq. \ref{eq:delaysuppress}, $\mathcal{D}(\mathcal{P}_{sc},L)$ in Eq. \ref{eq:oracle_constrain_distance} is fulfilled. Delay suppressed \textsc{Ormer} MedDS is given as Algorithm \ref{alg:ormer_2slot}.

\subsection{Slot Encoding}\label{subsec_slot}

Slot is the minimum storage unit of smart contract. In Ethereum, one slot has $256$ bits of space. As default, one variable would take one whole slot to be stored in smart contract, so encoding bits inside one slot is important for gas saving. \textit{Tick} introduced in \cite{uniswapv3} could encode a $p_t$ as a signed $24$-bits number with reasonable precision:

\begin{equation}\label{eq:tick}
\tau=\text{log}_{1.0001}p_t
\end{equation}

According to equation \ref{eq:tick}, the price movements of one tick can still be detected as a precision of 1 basis point (0.01\%), which would suit most of financial applications. With tick, we are able to store the five marker height information into $120$ bits. Another $24$ bits is allocated for storing the last estimation tick. Based on remaining bits in a slot, five marker positions, observation windows size, and observation count are limited as unsigned 16-bits integers, which are $65,535$ in maximum in digit. With slot encoding and assistance of bit computation tricks, $\mathcal{G}(\mathcal{P}_{sc},L)$ in Eq. \ref{eq:oracle_constrain_gas} is fulfilled. Additionally, for \textsc{Ormer} MedDS, there are two slots need to be encoded as illustrated in Algorithm \ref{alg:ormer_2slot}.

Detailed bits allocation for EVM compatible smart contract slot is shown in Table \ref{tab:slot}. It is worth mentioning that in EVM blockchains, writing pure zero on blockchain slot would cost extra gas fee. Thus, for state initialization in \textsc{Ormer}-SlidingWindow, stored values should be carefully manipulated in the smart contract slot. Frequently used values are also hard coded as constants in the contract, so the gas would only be consumed once during the contract publish process.

{
\setlength{\textfloatsep}{0pt}
\begin{algorithm}[t!]
    \caption{\textsc{Ormer} MedDS}\label{alg:ormer_2slot}
    \Input{Observation window $L$,\\Incoming data point $p_t$}
    \Output{Ormer estimation $\hat{p}_t^o$}
    
    \If{\text{not Initialized}} {
        $O_{\text{full}} \gets new\textsc{ OrmerMED(}T\textsc{)}$\;
        $O_{\text{half}} \gets new\textsc{ OrmerMED(}\lfloor{T/2}\rfloor\textsc{)}$\;
    }
    $O_{\text{full}}$.Update($p_t$)\;
    $O_{\text{half}}$.Update($p_t$)\;
    $\hat{p}_{\text{full}}\gets O_{\text{full}}.getMedian()$\;
    $\hat{p}_{\text{half}}\gets O_{\text{half}}.getMedian()$\;
    $\hat{p}_t^o \gets (\hat{p}_{\text{half}}/{\hat{p}_{\text{full}}}) \cdot (\hat{p}_{\text{half}}+\hat{p}_{\text{full}})/2$\;
\end{algorithm}
}

\begin{table}[]
\caption{Bits allocation details for one slot.}
\label{tab:slot}
\resizebox{\columnwidth}{!}{%
\begin{tabular}{@{}r|rcc@{}}
\toprule
\textbf{Category} & \textbf{Information} & \multicolumn{1}{r}{\textbf{Signed}} & \multicolumn{1}{r}{\textbf{Bits Allocated}} \\ \midrule
\multirow{3}{*}{Window Info} & Window Size & \ding{55} & 16 \\
 & Observation Count & \ding{55} & 16 \\
 & Last Median Estimation & \ding{51} & 24 \\ \midrule
\multirow{2}{*}{Price Marker} & Marker Positions & \ding{55} & $16\times5$ \\
 & Marker Heights & \ding{51} & $24\times5$ \\ \bottomrule
\end{tabular}%
}
\end{table}

\section{Evaluation}\label{sec_eval}

\subsection{Experiment Setup}
\textbf{Dataset.} We have collected $4,130,000$ spot prices from real-world swapping pools (DEXs) from 2022-01-13 to 2023-09-13 \cite{usdtweth}, and the original price feed is sampled according to Poisson distribution corresponds to random trading activities. $839,728$ of real-world spot prices of DEX are obtained after sampling. Besides, we have also collected $42,082,934$ (k-line in seconds) DEXs' corresponding market prices ranging from 2022-01-01 to 2023-08-31 from Binance (CEX), and $134,669$ off-chain oracle prices ranging from 2020-01-15 to 2023-07-27 from Chainlink (off-chain pricing oracle) \cite{breidenbach2021chainlink}. In our evaluations, we assume the CEX price feed (e.g., from Binance) reflects accurately about market price feed $\mathcal{P}_{\text{true}}$.

\textbf{Baselines and Comparisons}. There are two baselines selected respectively for quantifying the metrics and comparing the performances:
\begin{itemize}[left=0pt]
    \item\textbf{CEX Price Feed}: it is selected as the baseline for calculating the metrics in \S\ref{subsec_metrics}. The improved metrics indicate that the oracle price feed will be kept very close to the market price, so attackers need to spend more funds to deviate the price. In other words, there is less opportunity to initiate a successful attack on a price feed with low delay and deviation from market.
    \item\textbf{True Median}: it is selected as the baseline for comparing the performances of pricing oracles. 
\end{itemize}

We consider four (Chainlink, TWAP, EMA, and True Median) existing pricing oracles and implement three of them (TWAP, EMA, and True Median) as there are no available implementations can be tested\footnote{Uniswap's implementation of TWAP requires cross-contract invocations, and the gas consumption could be higher than the result shown in Table \ref{tab:cmp}.}. We record and compare their performances by feeding the above dataset in a streaming way.

\textbf{Implementations}. \textsc{Ormer} MED and \textsc{Ormer} MedDS are implemented as one \textsc{Ormer} Contract with Solidity language that is Ethereum Virtual Machine (EVM) compatible for public blockchains such as Ethereum, Polygon, and Avalanche. In order to further reduce gas consumption, considering processing ticks on-chain would involves complex exponent computation, our contract implementation employs the tick library developed by Uniswap V3 with magic number tricks that significantly reduce on-chain tick-related computations \cite{uniswapv3}. Moreover, since float number is not originally supported on EVM compatible blockchain, a gas efficient 64x64-bits fixed point number computing framework is used \cite{abdk}.

For simplicity on blockchain wallet management, besides evaluating on a public blockchain (Ethereum Sepolia), all the implemented pricing oracles are also evaluated on locally deployed EVM compatible blockchain network \cite{hardhat}. The blockchain full nodes are run on a workstation (Intel i9-13900K CPU, 96GB Memory) and several Nvidia Jetson Orins in a local network.

\subsection{Metrics}\label{subsec_metrics}

\begin{table*}[ht!]
\centering
\caption{Evaluation results on different price feeds with window size $L=25$ blocks compared to CEX price feed.}
\label{tab:cmp}
\setlength{\tabcolsep}{6pt}
\renewcommand{\arraystretch}{1.5}

\resizebox{\textwidth}{!}{%
\begin{tabular}{@{}r|rrrrrrrrr|rrr|rl|ll@{}}
\toprule
 & \multicolumn{9}{c|}{\textbf{Deviation Resistance}} & \multicolumn{3}{c|}{\textbf{Market Delay}} & \multicolumn{2}{c|}{\textbf{On-chain Cost}} & \multicolumn{2}{c}{} \\ \cmidrule(lr){2-15}
\multirow{-2}{*}{\textbf{Category}} & \textbf{Price Feed} & \textbf{MAE} & \textbf{MSE} & \textbf{MedAE} & \textbf{MaxErr} & \textbf{TDP} & \textbf{TDG} & \textbf{MAPE} & \multicolumn{1}{c|}{\cellcolor[HTML]{EFEFEF}\textbf{\begin{tabular}[c]{@{}c@{}}Stationary\\ Score\end{tabular}}} & \multicolumn{1}{c}{\textbf{\begin{tabular}[c]{@{}c@{}}Delay\\ (Window)\end{tabular}}} & \multicolumn{1}{c}{\textbf{\begin{tabular}[c]{@{}c@{}}Delay\\ (All)\end{tabular}}} & \multicolumn{1}{c|}{\cellcolor[HTML]{EFEFEF}\textbf{\begin{tabular}[c]{@{}c@{}}Delay\\ Score\end{tabular}}} & \multicolumn{1}{c}{\textbf{\begin{tabular}[c]{@{}c@{}}Gas\\ Consumption\end{tabular}}} & \multicolumn{1}{c|}{\cellcolor[HTML]{EFEFEF}\textbf{\begin{tabular}[c]{@{}c@{}}Gas\\ Score\end{tabular}}} & \multicolumn{2}{c}{\multirow{-2}{*}{\textbf{\begin{tabular}[c]{@{}c@{}}Overall Resistance Efficiency\\ Score\end{tabular}}}} \\ \midrule
\begin{tabular}[c]{@{}r@{}}Off-chain Oracle\\ (Centralized)\end{tabular} & Chainlink & 3.342 & 24.605 & 2.276 & 135.589 & 1.166 & 0.006 & 0.178 & \cellcolor[HTML]{EFEFEF}0.003 & 313.2 & 249 & \cellcolor[HTML]{EFEFEF}2.736 & N/A & \multicolumn{1}{r|}{\cellcolor[HTML]{EFEFEF}N/A} & \multicolumn{2}{c}{N/A} \\ \midrule
 & TWAP & 4.651 & 50.484 & 3.122 & 202.341 & 2.540 & 0.014 & 0.253 & \cellcolor[HTML]{EFEFEF}0.109 & 627.8 & 1,049 & \cellcolor[HTML]{EFEFEF}0.917 & 199,474 & \cellcolor[HTML]{EFEFEF}1.794 & \cellcolor[HTML]{EFEFEF}1.106 & \scorebar{110.63} \\
 & EMA & 4.255 & 40.545 & 2.948 & 184.079 & 2.028 & 0.011 & 0.231 & \cellcolor[HTML]{EFEFEF}0.013 & 404.7 & 830 & \cellcolor[HTML]{EFEFEF}1.246 & 213,981 & \cellcolor[HTML]{EFEFEF}1.673 & \cellcolor[HTML]{EFEFEF}1.170 & \scorebar{117.00} \\
 & True Median & 4.735 & 55.437 & 3.091 & 204.251 & 2.798 & 0.016 & 0.257 & \cellcolor[HTML]{EFEFEF}1.000 & 555.0 & 983 & \cellcolor[HTML]{EFEFEF}1.000 & 357,909 & \cellcolor[HTML]{EFEFEF}1.000 & \cellcolor[HTML]{EFEFEF}1.000 & \scorebar{100.00} \\ \cmidrule(l){2-17} 
 & \textbf{\textsc{Ormer} MED} & \textbf{4.867} & \textbf{57.334} & \textbf{3.192} & \textbf{225.891} & \textbf{2.883} & \textbf{0.016} & \textbf{0.264} & \cellcolor[HTML]{EFEFEF}\textbf{0.454} & \textbf{555.2} & \textbf{1,162} & \cellcolor[HTML]{EFEFEF}\textbf{0.896} & \textbf{169,062} & \cellcolor[HTML]{EFEFEF}\textbf{2.117} & \cellcolor[HTML]{EFEFEF}\textbf{1.296} & \scorebar{129.59} \\
\multirow{-5}{*}{\begin{tabular}[c]{@{}r@{}}On-chain Oracle\\ (Decentralized)\end{tabular}} & \textbf{\textsc{Ormer} MedDS} & \textbf{3.940} & \textbf{33.412} & \textbf{2.801} & \textbf{168.874} & \textbf{1.677} & \textbf{0.009} & \textbf{0.214} & \cellcolor[HTML]{EFEFEF}\textbf{0.006} & \textbf{325.7} & \textbf{532} & \cellcolor[HTML]{EFEFEF}\textbf{1.793} & \textbf{284,317} & \cellcolor[HTML]{EFEFEF}\textbf{1.259} & \cellcolor[HTML]{EFEFEF}\textbf{1.222} & \scorebar{122.20} \\ \bottomrule

\multicolumn{17}{p{1.5\textwidth}}{\footnotesize \textit{Note}: All the metrics with white background are regarded as better if they are closed to 0. All the scores with gray background are regarded as better if it is higher. MAE: Mean Absolute Error, MSE: Mean Squared Error, MedAE: Median Absolute Error, MaxErr: Max Error, TDP: Tweedie Deviance Error with power of $1$ (Poisson distribution), TDG: Tweedie Deviance Error with power of $2$ (Gamma distribution), MAPE: Mean Absolute Percentage Error. The unit of MAPE is percentage, the unit of TDP is in $10^{-2}$ USD (\$), the unit of TDG is in $10^{-3}$ USD (\$), the unit of Delays are second (s), the unit of Gas Consumption is in GWei, Scores (Stationary, Delay, Gas, Resistance Efficiency) do not have units, and all the other metrics are in USD (\$). Calculation of Stationary Score discards the units of TDP and TDG.}

\end{tabular}
}
\end{table*}

According to our knowledge, metrics evaluating the performance of a blockchain pricing oracle is missing in the literature. In this paper, we evaluate the performance of price feed constructed by blockchain pricing oracles from three aspects: deviation resistance, delay and on-chain cost. Among which, evaluation on deviation resistance and delay corresponds to $\mathcal{D}(\mathcal{P}_{sc},L)$ described in Eq. \ref{eq:oracle_constrain_distance} (i.e., manipulation resistance), while evaluation on on-chain cost corresponds to $\mathcal{G}(\mathcal{P}_{sc},L)$ described in Eq. \ref{eq:oracle_constrain_gas} (i.e., gas efficiency). 

Based on these metrics, we further calculate three scores: Stationary Score, Delay Score, and Gas Score, which use True Median as the baseline to eventually compare their effectiveness a unified way. True Median is sleeted as it is the most promising manipulation-resistant solution besides \textsc{Ormer}. In the end, weighted average based \textbf{Resistance Efficiency Scores}, that jointly consider three scores, are first proposed for overall evaluation for on-chain pricing oracles. The calculation of Resistance Efficiency Scores can be used in future researches, filling the absence of a comparison metric.

\textbf{Deviation Resistance}.

Besides the commonly used sub-metrics like MAE, MSE, we also consider Tweedie deviance error. Tweedie deviance error is a regression result evaluation metric that elicits predicted expectation values of regression targets. We employed special cases of Tweedie deviance error formula as our evaluation metrics. The cases of listed $TD_p(y, \hat{y})$ are equivalent to Mean Square Error (MSE), Mean Poisson Deviance, and Mean Gamma Deviance respectively, which would give us comprehensive intuitions regarding the correctness of the constructed feed with potential flexible variance structure.

Particularly, if $\hat{y}_i$ from price feed $\hat{y}$ is the predicted value of the $i$-th sample in observation window with number of $n$ samples, and $y_i$ from price feed $y$ is the corresponding baseline value, then the Tweedie deviance error formula $TD(y, \hat{y})$ is given by \cite{scikit-learn}:

\begin{equation}
\resizebox{\linewidth}{!}{$
TD_p(y, \hat{y}) = \frac{1}{n} \sum_{i=0}^{n - 1}
\begin{cases}
(y_i-\hat{y}_i)^2, & \text{ ($p=0$, Normal)}\\
2(y_i \log(y_i/\hat{y}_i) + \hat{y}_i - y_i),  & \text{ ($p=1$, Poisson)}\\
2(\log(\hat{y}_i/y_i) + y_i/\hat{y}_i - 1),  & \text{ ($p=2$, Gamma)}
\end{cases}
$}
\end{equation}

With this, we further define \textbf{Stationary Score} $\text{Scr}_{X}^{\text{St}}$ to comprehensively reflects the resistance of a price feed by considering three cases of Tweedie deviance at the same time:

\begin{equation}
3(\frac{1}{\text{Scr}_{X}^{\text{St}}}-1)=\sum_{p=0}^{2}\left[ TD_p(\text{CEX},X)-TD_p(\text{CEX},\text{MED})\right] ^2
\end{equation}

\noindent where $X$ represents the price feed under evaluation, CEX refers to $\mathcal{P}_\text{true}$, and MED refers to price feed provided by True Median. $\text{Scr}_{X}^{\text{St}}$ reflects how similar the $X$ performs like True Median, and a higher score indicates a more stationary price feed. There are overlaps among the regression metrics we have evaluated, so only three cases of Tweedie deviance are selected as the score definition, since they could cover the most concerned aspects of the regression evaluation. Notably, a low $\text{Scr}_{X}^{\text{St}}$ does not mean the price feed is vulnerable in practice (e.g., Chainlink). It need to be further balanced with the delay metrics to represent our manipulation-resistant goal.

\textbf{Delay}. We use cross correlation to evaluate the time delay in our experiments. Time delay of price feed $Y$ comparing to price feed $X$ is measured according to:

\begin{equation}\label{eq:crosscorr}
\text{Delay}(X,Y, \mathcal{T}) = \mathop{max}\limits_{\mathop{t\in \mathcal{T}}} \frac{\mathbb{E}[(X - \mu_{X})(Z - \mu_Z)]}{\sigma_{X}\sigma_{Z}}
\end{equation}

\noindent where $\mu$ and $\sigma$ are corresponding mean and variance to price feeds, $Z=\{(t_Z,p_Z)|t_Z\in\{{t_Y}_1-t,\dots,{t_Y}_n-t\},p_Z\in\{{p_Y}_1,\dots,{p_Y}_n\}\}$. Price feed $X$ is fixed to CEX price feed in our experiment setup on overall delay, while $\mathcal{T}$ is capped with $1800$s.

We also define the \textbf{Delay Score} $\text{Scr}_{X}^{\text{De}}$ to represent how good the price feed reflects the delay more intuitively:

\begin{equation}
{\text{Scr}_{X}^{\text{De}}}=\frac
{\text{Delay(CEX, MED, $\mathcal{T}_\text{all}$)} + \text{Delay(CEX, MED, $\mathcal{T}_\text{win}$)}}
{\text{Delay(CEX, $X$, $\mathcal{T}_\text{all}$)} + \text{Delay(CEX, $X$, $\mathcal{T}_\text{win}$)}}
\end{equation}

\textbf{On-chain Cost}.
The on-chain cost is given by the gas consumption of an on-chain smart contract invocation, which is measured by recording the blockchain transaction commit. To invoke smart contract functions through blockchain transactions, the DeFi participants need to pay the fee for code executions (i.e., Gas Fee). The fee paid to the block builder can be directly obtained from the receipts of transactions, and can be further calculated as gas scores. We define the \textbf{Gas Score} $\text{Scr}_X^{\text{Gas}}$ by comparing to True Median that needs to record full historical data (i.e., high gas consumption):

\begin{equation}
{\text{Scr}_X^{\text{Gas}}}=\frac{\text{Gas}_{\text{MED}}}{\text{Gas}_X}
\end{equation}

\textbf{Overall Resistance Efficiency}.
In order to evaluate \textsc{Ormer} Contract by jointly considering deviation resistance, delay, and on-chain cost, we define \textbf{Resistance Efficiency Score} by taking weighted average of the Stationary Score, Delay Score, and Gas Score:

\begin{equation}
\text{Scr}_X^{\text{RE}}=(\omega_0\text{Scr}_{X}^{\text{St}}+\omega_1\text{Scr}_{X}^{\text{De}}+\omega_2\text{Scr}_X^{\text{Gas}})/\sum_{i=0}^{2}\omega_i
\end{equation}

\noindent where weights vector $[\omega_0,\omega_1,\omega_2]$ is given by $[1,2,2]$ for paying more attention to the challenges (i.e., delay and on-chain cost) we mainly focus on in this work.

\subsection{Effectiveness on Manipulation Resistance}
The manipulation resistances of price feeds constructed by existing oracles and \textsc{Ormer} are evaluated by jointly considering three aspects: regression metrics comparing to CEX, stationary scores, and delay score.

The regression metrics are used to quantify the similarity of the price feed comparing to the external market price (e.g., CEX). As we mentioned above, if an on-chain price feed is constructed more similar to the external source (i.e., regression metrics are smaller), the cost on deviating the price will be raised. Thus, there is less opportunity to proceed a price manipulation attack on a price feed with low delay and deviation from market. The results in Table \ref{tab:cmp} illustrate that \textsc{Ormer} MedDS outperforms all the implemented on-chain oracle in our setup with real-world trading data. Surprisingly, decentralized on-chain \textsc{Ormer} MedDS only shares $17.9\%$ more MAE comparing to a centralized off-chain oracle, which is well-known as a reliable external data source. While \textsc{Ormer} MED trades for more stationary price feed with $4.6\%$ more MAE comparing to the SotA TWAP oracle.

The stationary score is used to reflect the similarity between a price feed and the True Median, which is derived from regression metrics to the external market price. A higher score indicates that the price feed is less sensitive to an outlier price, which is potentially manipulated by malicious parties. The stationary scores in Table \ref{tab:cmp} clearly distinguish \textsc{Ormer} MED from other price feeds (e.g., $416.5\%$ to TWAP), indicating \textsc{Ormer} MED's significant performance in estimating the True Median in a streaming way with low gas consumption. \textsc{Ormer} MedDS's score is lower because it is translated to fit the latest market price after estimating the True Median. However, this high idleness to fluctuation would result in high market delay, which may expose extra attack interfaces \cite{heimbach2024non,wang2022cyclic}. So the stationary scores are further balanced with the following delay scores to measure the effectiveness on manipulation resistances.

The delay score reflects the lag between oracle price feed and external market price comparing to True Median. A higher score indicates a better tracing to the truthful market dynamics, and an abnormal fluctuation caused by outlier prices would recover faster. Surprisingly, with the window size set to $300$s, the delay measured for \textsc{Ormer} MedDS reaches $532$s, only $50.7\%$ of the SotA TWAP. The result indicates that \textsc{Ormer} MedDS may have the potential for dealing with price feed prediction tasks. If the delay is estimated in 1h-window, the statistical result of \textsc{Ormer} MedDS reaches $(325.7\pm891.7)$s, even comparable to off-chain oracle Chainlink with $(313.2\pm872.1)$s. The results of $\text{Scr}_{X}^{\text{De}}$ illustrate that \textsc{Ormer} MedDS outperforms TWAP and True Median for $95.5\%$ and $79.3\%$ respectively. Based on the results, we can conclude that the delay suppression method introduced in \S\ref{subsec_delay} works well on real-world dataset. For \textsc{Ormer} MED, although less sensitive to market dynamics by design, still achieves more than $97.7\%$ performance of TWAP according to Delay Score.

Jointly considering the aspects mentioned by taking average of Station Scores and Delay Scores, \textsc{Ormer} MedDS achieves $90.0\%$ performance of True Median and $65.7\%$ performance of Chainlink regarding deviation resistance, with $75.4\%$ improvement to TWAP; \textsc{Ormer} MED achieves $67.5\%$ performance of True Median and $49.3\%$ performance of Chainlink, with $31.6\%$ improvement to TWAP. Therefore, \textsc{Ormer}  is manipulation resistant.

\begin{table}[]
\caption{Statistical results of gas consumption and the partitioning percentages of $Update()$ and $Query()$.}
\setlength{\tabcolsep}{4.2pt}
\renewcommand{\arraystretch}{1.3}
\label{tab:gas_detail}
\resizebox{\columnwidth}{!}{%
\begin{tabular}{@{}r|clcl@{}}
\toprule
\multirow{2}{*}{\textbf{Price Feed}} & \multicolumn{4}{c}{\textbf{Invocation Gas Cost (GWei)}} \\ \cmidrule(l){2-5} 
 & \textbf{Update()} & \multicolumn{1}{l|}{\textbf{Percentage}} & \textbf{Query()} & \textbf{Percentage} \\ \midrule
TWAP & 99,614±1 & \multicolumn{1}{l|}{49.94\%} & 99,860±7,629 & 50.06\% \\
EMA & 120,733±1,600 & \multicolumn{1}{l|}{56.42\%} & 93,248±7,740 & 43.58\% \\
True Median & 45,700±4,433 & \multicolumn{1}{l|}{12.77\%} & 312,209±29,175 & 87.23\% \\ \midrule
\textbf{Ormer MED} & \textbf{124,183±34,121} & \multicolumn{1}{l|}{\textbf{73.45\%}} & \textbf{44,879±589} & \textbf{26.55\%} \\
\textbf{Ormer MedDS} & \textbf{214,000±54,980} & \multicolumn{1}{l|}{\textbf{75.27\%}} & \textbf{70,317±1,075} & \textbf{24.73\%} \\ \bottomrule
\end{tabular}%
}
\end{table}

\begin{figure*}[ht!]
\centering
\includegraphics[width=0.99\textwidth]{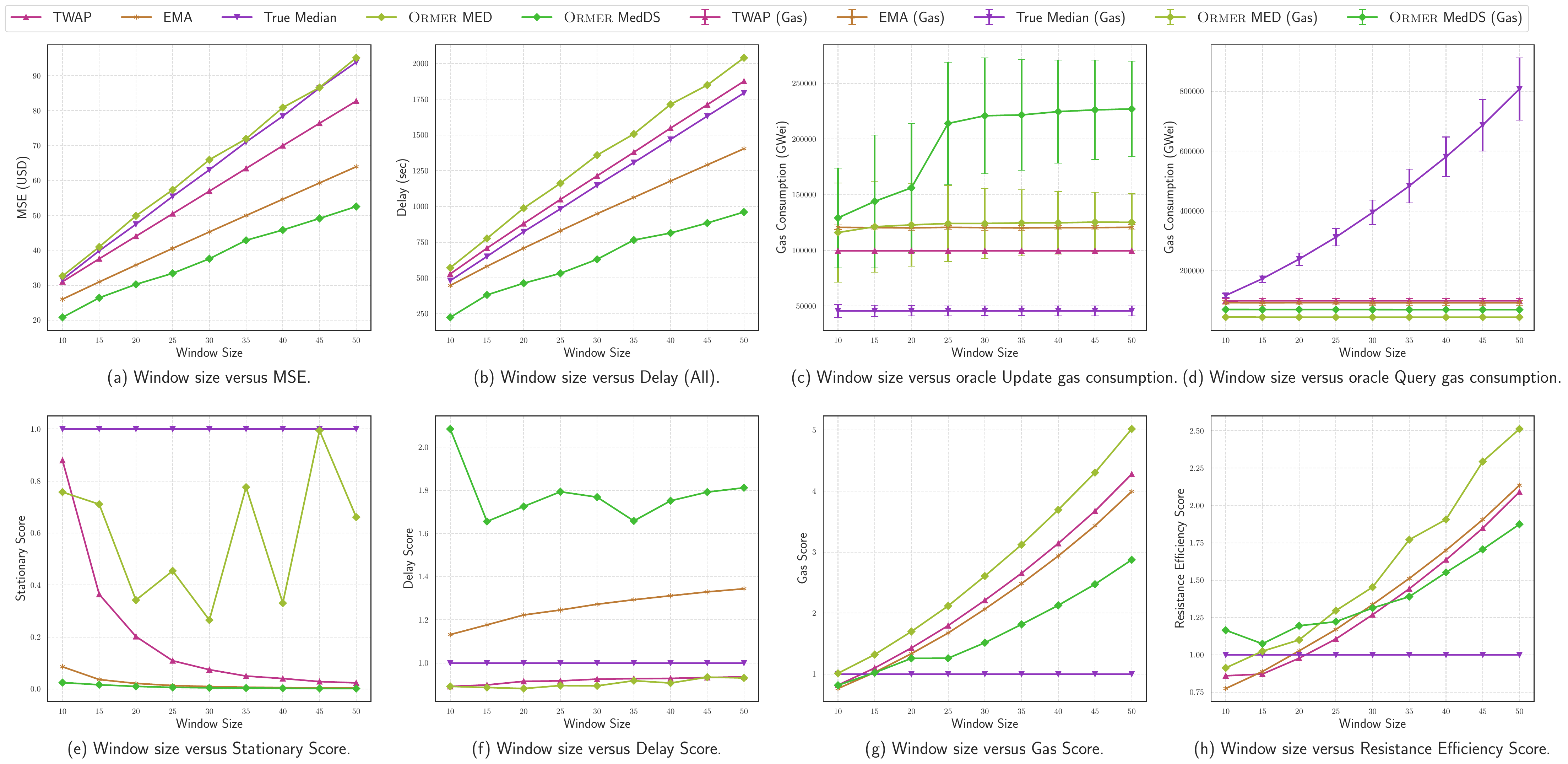}
\caption{Ablation studies on influences of window size on evaluation metrics.}
\label{fig:ablation}
\end{figure*}

\subsection{Effectiveness on Gas Efficiency}
In our experiment setup, the gas is measured on $10,000$ contract invocation transactions from two separate blockchain transaction commits: invocation of $Update()$ function that triggers state update of oracle contract, and invocation of $Query()$ function that simply returns current $p_t^o$. In real-world scenario, $Update()$ would be triggered by traders who swap cryptocurrencies in the exchanges, while $Query()$ would be invoked by downstream DeFi applications requiring reliable price feed.

Gas score results in Table \ref{tab:cmp} indicate that \textsc{Ormer} MED outperforms True Median for $111.7\%$. It even outperforms TWAP for $18.0\%$, which is widely deployed in production as its low gas consumption. As for \textsc{Ormer} MedDS, since it requires one more auxiliary storage slot for delay suppression (i.e., more manipualtion resistance), it only outperforms True Median for $25.9\%$. But it still reaches $70.2\%$ performance of SotA TWAP regarding gas efficiency.

As shown in Table \ref{tab:gas_detail}, comparing to other existing on-chain oracles, \textsc{Ormer} Contract spends more cost on $Update()$ invocations, while True Median spends most of the cost on $Query()$. This is because \textsc{Ormer} algorithms require complex state updates by writing more information on blockchain (writing data on-chain is expensive), but querying estimated median would require less computations. For True Median, to reduce gas cost, we maintain a pre-allocated space in our implementation. That is to say, updating states involves minimum slot storage change, but querying would require more reading-extensive operations to rebuild price feed with given window size.

As for detailed comparisons to TWAP, \textsc{Ormer} MedDS with higher gas consumption would cost $(70,317\pm1,075)$ GWei (around $21.1$ USD, estimated according to: $3,000$ USD/ETH, $100$ Gas Price) for querying oracle price, even less that TWAP oracle querying historical data in ring buffer with $(99,860\pm7,629)$ GWei (around $30.0$ USD) \cite{uniswapv3}. This would significantly help to reduce trading cost and elevate liquidity efficiency to the DeFi market. \textsc{Ormer} MED even reaches $44.9\%$ gas consumption on $Query()$ compared to TWAP. For $Update()$ that requires more complex smart contract storage changing, \textsc{Ormer} MED and \textsc{Ormer} MedDS cost $24.7\%$ and $114.8\%$ more gas respectively for more manipulation resistances.

To conclude with \textbf{Resistance Efficiency Score} that jointly considers manipulation resistance and gas efficiency, as shown in Table \ref{tab:cmp}, both modes of \textsc{Ormer} Contract outperform price feeds constructed by existing pricing oracles with up to $29.6\%$ improvement comparing to True Median. Based on the obtained evaluation results, we can conclude that the \textsc{Ormer} Contract implementation is manipulation-resistant and gas-efficient.

\subsection{Ablation Study}
To better illustrate the effectiveness of \textsc{Ormer} Contract, ablation studies on window size choices are conducted on five on-chain oracles (i.e., TWAP, EMA, True Median, \textsc{Ormer} MED, and \textsc{Ormer} MedDS).

As shown in Figure \ref{fig:ablation}(a) and Figure \ref{fig:ablation}(b), MSE and Delay (All) would increase lineally when window size grows. For gas consumption, there are no significant changes in updating oracles except \textsc{Ormer} MedDS as shown in Figure \ref{fig:ablation}(c), since \textsc{Ormer} algorithms require to gather sufficient markers to boot. To be more specific, cost of \textsc{Ormer} MedDS would surge up $30\%$ when window size is changed from $23$ to $24$; for \textsc{Ormer} MED, a smaller surge occurs at window size $9$. For comparison fairness, we choose to set the number larger than $23$ in Table \ref{tab:cmp} (25 is randomly selected, which is also widely used in DeFi according to the literature and our observations). For querying cost in Figure \ref{fig:ablation}(d), there are no significant changes, except True Median as it would require more slot readings to find the accurate median with a bigger window size.

Stationary scores of TWAP, EMA, and \textsc{Ormer} MedDS would decrease exponentially as window size grows, as illustrated in Figure \ref{fig:ablation}(e), since the shape would be distorted significantly from CEX compared to True Median. Meanwhile, \textsc{Ormer} MED demonstrates its great estimation of True Median with best Stationary Score among all the evaluated oracles except window size with $10$. Figure \ref{fig:ablation}(f) demonstrates the effectiveness of the Delay Suppression method proposed in \S\ref{subsec_delay}, while Figure \ref{fig:ablation}(g) illustrates the effectiveness of the Slot Encoding method proposed in \S\ref{subsec_slot}. In the end, Figure \ref{fig:ablation}(h) demonstrates the overall resistance efficiency of \textsc{Ormer}.

As the only tunable parameter, in this paper, we yield downstream applications to choose the number according to their practical businesses, so the optimal number is not pursued in our work (e.g., smaller for low-delay derivative trading, greater for long-term trend analysis). Dynamically tuning the number based on market conditions would be a great direction for future works. For this work, we would recommend using a fixed small window size to track up-to-date market price.

\begin{figure}[t!]
\centering
\includegraphics[width=0.42\textwidth]{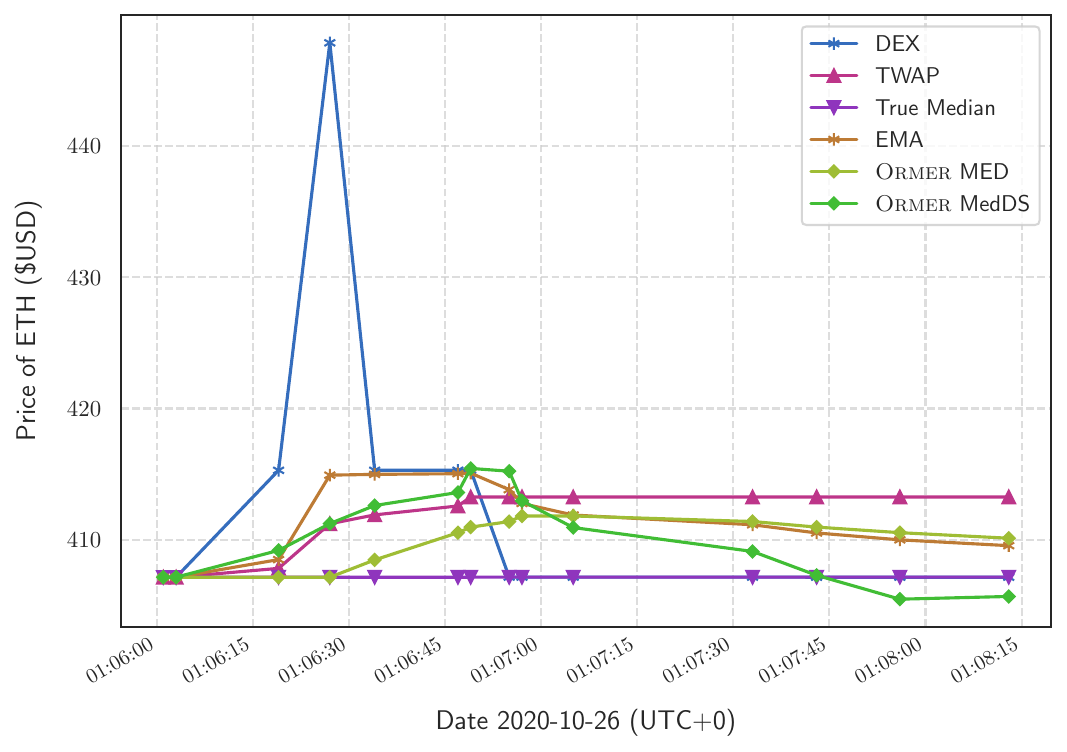}
\caption{Case Study of manipulation resistance on real-world token swapping historical data \cite{eminence2020attack}.}
\label{fig:case_study}
\end{figure}

\subsection{Case Study on Manipulation Attacks}
We have shown the significances of both \textsc{Ormer} MED and \textsc{Ormer} MedDS in a real-world flash loan attack in Figure \ref{fig:ormer}. To make manipulation-resistant property more clear, we provide a zoom-in case study on clipped real-world trading data as shown in Figure \ref{fig:case_study}. The original DEX price datapoints in the clipped period of time remains small fluctuations around $407$ USD. We replayed a manipulation attack on $3$-th, $4$-th, $5$-th, $6$-th, and $7$-th datapoints. TWAP, True Median, EMA, \textsc{Ormer} MED, and \textsc{Ormer} MedDS are then all employed with window size of $12$ for pricing oracle feed constructions. The results intuitively confirms the stationary property of \textsc{Ormer} MED and True Median as they stay inactive for the peak, while both \textsc{Ormer} MedDS and EMA are sensitive to the changes. However, EMA is misled by the $4$-th peak in the following prices, while \textsc{Ormer} MedDS stays stationary at $4$-th peak and self-corrects the price quickly after 01:07:00 based on the latest market dynamics. Both \textsc{Ormer} MED and EMA demonstrates some sort of self-correction property after manipulation, while TWAP behaves lagged in recovering. To conclude, all the pricing oracles shown in Figure \ref{fig:case_study} demonstrates different levels of manipulation resistance, but only \textsc{Ormer} MedDS follows latest DEX price feed actively.

To conclude, \textsc{Ormer} MED suits for use cases that are more sensitive to gas (e.g., on-chain order book \cite{Synfutures}), while \textsc{Ormer} MedDS would be a promising alternative to the SotA TWAP as a new pricing oracle infrastructure. Both oracles provide a more cost-efficient option by balancing all the metrics.

\section{Discussion}
\textbf{Potential Risks.} According to Figure \ref{fig:marker}, we would prefer to update current marker at index $i$ with its two neighbor markers ($i-1$ and $i+1$) due to the gas consideration. This constrains the number of coefficients to limited $3$: $a$, $b$, and $c$. In fact, estimating median through piecewise-parabolic formula with order of $2$ would also deviates when dealing with tremendous extreme cases, saying $10,000\%$ of price deviation. To better accommodate such cases, higher order polynomial formula is needed (e.g., $h_i=an_i^4+bn_i^3+cn_i^2+dn_i+e$). However, to solve the higher order equations, more linear combination computations would be needed for the estimation of single update of $h_i'$. A possible solution is to fix the coefficient limitation while trying to solve formula like $h_i=an_i^6+bn_i^2+c$. But our evaluation results indicate that the algorithm would also be impractical for on-chain pricing oracle regarding gas consumption. That is to say, the reason that piecewise-parabolic formula is selected as our core estimation formula in \textsc{Ormer} is based on the trade-off between robustness and gas consumption.

The Delay Suppression introduced in \S\ref{subsec_delay} may introduce new risks of inaccurate price predictions. However, the threats can be negligible in practice under the assumptions in this work. First, according to large-scale real-would evaluation results, which potentially contains unknown attacks, the overall evaluations on such dataset proves the soundness of \textsc{Ormer}'s manipulation-resistant property. Second, the cost-of-attacks on SotA oracles are significantly lower than \textsc{Ormer} theoretically. As an example, a manipulation on a single block (i.e., time step) is sufficient to influence the output of TWAP, while an observable attack on \textsc{Ormer} requires successive multi-block manipulations, which is possible but very hard to achieve and shares negligible possibility in real-world scenarios.

\textbf{Scalability.} As \textsc{Ormer} Contract is implemented with Solidity, which could be directly migrated to EVM compatible blockchains without any modification (e.g., Tron, Polygon). For other chains (e.g., Solana), it would require rewriting smart contracts with different programming languages (e.g., Rust). Therefore, the deployment could be trivial.

\section{Related Work}

\noindent \textbf{Attack Detection}. One possible countermeasure is to detect and deny transactions containing unlawful operations online (e.g., flash loan based manipulations) before they are confirmed on-chain. Researches have developed automated pattern detection methods for attack identification \cite{xia2023detecting,chen2024flashsyn,deng2024safeguarding,xi2024pomabuster}. However, detecting such patterns according to semantic analysis requires additional domain knowledge and complicated execution logic on smart contracts (e.g., smart contract fuzzing \cite{qian2024mufuzz}), which is thereby hard to be implemented and deployed into DeFi applications. There are also studies focusing on analyzing offline transaction histories to find out fraudulent users, which can be further blocked by the oracle maintainers \cite{zhong2024bitlink,elmougy2023demystifying}. But maintaining a blacklist on-chain would cost a huge amount of money, and querying a growing list would introduce new problems for on-chain smart contracts.

\noindent \textbf{Blockchain Pricing Oracle}. Another countermeasure is mitigating the manipulation impacts through a pricing oracle. Currently, TWAP is widely used both in Decentralized Finance (DeFi) and Traditional Finance (TradFi) providing reference price data feed that smooth the original price data feed incoming from an upstream Price Data Source (e.g., DEX, Stock Exchange). The goal of such mechanism is to provide a relatively stable pricing data feed in an ever changing market, while providing fair resistance to the manipulation attack, even in TradFi. The underlying principle of TWAP is a sliding-window based moving average filter that smooths the price feed from the source. For the on-chain TWAP Pricing Oracle, in order to be more cost effective (i.e., gas efficient), only the accumulative price $a_t=\sum_{i=1}^{t}p_i$ of spot prices is updated and saved in the oracle contract once a new block is proposed at time $t$, and the current oracle reference price $p_t^o$ with observation window $L$ is estimated according to $p_t^o=\frac{\sum_{i=t-t_L}^t p_i}{t-t_{L}}=\frac{a_t-a_{t_L}}{t-t_L}$.

However, recent researches indicate that TWAP is not manipulation-resistant as expected \cite{mackinga2022twap,bentley2022manipulating,qin2021attacking}. Park et al. proposed a Kalman Filter based conformal prediction oracle contract that could give an uncertain price interval of current spot price \cite{park2023acon2}. But the reliability of the results constructed by their method relies on combining information from multiple data sources, while the pick of exact numerical current price estimation is not discussed in the paper. Another off-chain Price Data Source called Chainlink periodically queries data from multiple off-chain CEXs while constructing price feeds based on them, which are further relayed to on-chain contracts \cite{breidenbach2021chainlink}. Despite a price feed provided by Chainlink is efficient and reliable, it works in a fully off-chain manner, which is infeasible to be directly invoked by on-chain smart contracts, and centralized managed relaying contracts may pose potential trust issues. There are also some researches focus on outsourcing complex on-chain contract computation to trusted off-chain compute nodes, such as: \textsc{SMART} \cite{huang2024advancing}, POSE \cite{frassetto2023pose}, and Arbitrum \cite{kalodner2018arbitrum}. However, these methods would require introducing additional security assumptions to the system (e.g., Trusted Execution Environment) \cite{fang2022seframe,li2019scalable}.

\section{Conclusion}
In this paper, we propose an on-chain streaming median estimation algorithm called \textsc{Ormer}, which is implemented as \textsc{Ormer} Contract with two price feed mode: \textsc{Ormer} MED and \textsc{Ormer} MedDS. The evaluation on large amount of real-world data illustrates that \textsc{Ormer} Contract, as a pricing oracle, is manipulation-resistant and gas-efficient, which reaches significant result compared to existing methods. It is worth highlighting that querying on \textsc{Ormer} Contract consumes comparable gas to TWAP, while reaches same level of time delay in 1h-window setting to an off-chain oracle in a purely on-chain manner.

\bibliographystyle{IEEEtran}
\nocite{*}
\bibliography{IEEEabrv,ref}

\begin{thebibliography}{10}
\providecommand{\url}[1]{#1}
\csname url@samestyle\endcsname
\providecommand{\newblock}{\relax}
\providecommand{\bibinfo}[2]{#2}
\providecommand{\BIBentrySTDinterwordspacing}{\spaceskip=0pt\relax}
\providecommand{\BIBentryALTinterwordstretchfactor}{4}
\providecommand{\BIBentryALTinterwordspacing}{\spaceskip=\fontdimen2\font plus
\BIBentryALTinterwordstretchfactor\fontdimen3\font minus \fontdimen4\font\relax}
\providecommand{\BIBforeignlanguage}[2]{{%
\expandafter\ifx\csname l@#1\endcsname\relax
\typeout{** WARNING: IEEEtran.bst: No hyphenation pattern has been}%
\typeout{** loaded for the language `#1'. Using the pattern for}%
\typeout{** the default language instead.}%
\else
\language=\csname l@#1\endcsname
\fi
#2}}
\providecommand{\BIBdecl}{\relax}
\BIBdecl

\bibitem{eminence2020attack}
Etherscan, ``Eminence attack transaction,'' \url{https://etherscan.io/tx/0x35f8d2f572fceaac9288e5d462117850ef2694786992a8c3f6d02612277b0877}, 2020.

\bibitem{werner2022sok}
S.~Werner, D.~Perez, L.~Gudgeon, A.~Klages-Mundt, D.~Harz, and W.~Knottenbelt, ``Sok: Decentralized finance (defi),'' in \emph{Proceedings of the 4th ACM Conference on Advances in Financial Technologies}, 2022, pp. 30--46.

\bibitem{uniswapv3}
H.~Adams, N.~Zinsmeister, M.~Salem, R.~Keefer, and D.~Robinson, ``Uniswap v3 core,'' \emph{Tech. rep., Uniswap, Tech. Rep.}, 2021.

\bibitem{heimbach2022exploring}
L.~Heimbach, E.~Schertenleib, and R.~Wattenhofer, ``Exploring price accuracy on uniswap v3 in times of distress,'' in \emph{Proceedings of the 2022 ACM CCS Workshop on Decentralized Finance and Security}, 2022, pp. 47--53.

\bibitem{Aave}
A.~Labs, ``Aave protocol,'' \url{https://github.com/aave/aave-protocol/}, 2020.

\bibitem{Compound}
G.~H. Robert~Leshner, ``Compound: The money market protocol,'' \url{https://compound.finance/documents/Compound.Whitepaper.pdf}, 2019.

\bibitem{Synfutures}
S.~Team, ``Synfutures v3: The oyster amm model for next-gen defi derivatives [draft],'' \url{https://www.synfutures.com/v3-whitepaper.pdf}, 2024.

\bibitem{DefiLlama}
DefiLlama, ``Defillama,'' \url{https://defillama.com/}, 2024.

\bibitem{fritsch2021concentrated}
R.~Fritsch, ``Concentrated liquidity in automated market makers,'' in \emph{Proceedings of the 2021 ACM CCS Workshop on Decentralized Finance and Security}, 2021, pp. 15--20.

\bibitem{ni2024money}
W.~Ni, Z.~Yiwei, W.~Sun, L.~Chen, P.~Cheng, C.~J. Zhang, and X.~Lin, ``Money never sleeps: Maximizing liquidity mining yields in decentralized finance,'' in \emph{Proceedings of the 30th ACM SIGKDD Conference on Knowledge Discovery and Data Mining}, 2024, pp. 2248--2259.

\bibitem{mackinga2022twap}
T.~Mackinga, T.~Nadahalli, and R.~Wattenhofer, ``Twap oracle attacks: Easier done than said?'' in \emph{2022 IEEE International Conference on Blockchain and Cryptocurrency (ICBC)}.\hskip 1em plus 0.5em minus 0.4em\relax IEEE, 2022, pp. 1--8.

\bibitem{koutsoupias2019blockchain}
E.~Koutsoupias, P.~Lazos, F.~Ogunlana, and P.~Serafino, ``Blockchain mining games with pay forward,'' in \emph{The World Wide Web Conference}, 2019, pp. 917--927.

\bibitem{heimbach2024non}
L.~Heimbach, V.~Pahari, and E.~Schertenleib, ``Non-atomic arbitrage in decentralized finance,'' in \emph{2024 IEEE Symposium on Security and Privacy (SP)}.\hskip 1em plus 0.5em minus 0.4em\relax IEEE Computer Society, 2024, pp. 224--224.

\bibitem{wang2022cyclic}
Y.~Wang, Y.~Chen, H.~Wu, L.~Zhou, S.~Deng, and R.~Wattenhofer, ``Cyclic arbitrage in decentralized exchanges,'' in \emph{Companion Proceedings of the Web Conference 2022}, 2022, pp. 12--19.

\bibitem{mckay2022defi}
J.~McKay, ``Defi-ing cyber attacks,'' \url{https://tellingstorieswithdata.com/inputs/pdfs/final_paper-2022-jack_mckay.pdf}, 2022.

\bibitem{zhou2023sok}
L.~Zhou, X.~Xiong, J.~Ernstberger, S.~Chaliasos, Z.~Wang, Y.~Wang, K.~Qin, R.~Wattenhofer, D.~Song, and A.~Gervais, ``Sok: Decentralized finance (defi) attacks,'' in \emph{2023 IEEE Symposium on Security and Privacy (SP)}.\hskip 1em plus 0.5em minus 0.4em\relax IEEE, 2023, pp. 2444--2461.

\bibitem{kulkarni2023routing}
K.~Kulkarni, T.~Diamandis, and T.~Chitra, ``Routing mev in constant function market makers,'' in \emph{International Conference on Web and Internet Economics}.\hskip 1em plus 0.5em minus 0.4em\relax Springer, 2023, pp. 456--473.

\bibitem{xi2024pomabuster}
R.~Xi, Z.~Wang, and K.~Pattabiraman, ``Pomabuster: Detecting price oracle manipulation attacks in decentralized finance,'' in \emph{2024 IEEE Symposium on Security and Privacy (SP)}.\hskip 1em plus 0.5em minus 0.4em\relax IEEE, 2024, pp. 3923--3942.

\bibitem{qin2021attacking}
K.~Qin, L.~Zhou, B.~Livshits, and A.~Gervais, ``Attacking the defi ecosystem with flash loans for fun and profit,'' in \emph{International conference on financial cryptography and data security}.\hskip 1em plus 0.5em minus 0.4em\relax Springer, 2021, pp. 3--32.

\bibitem{lee2020measurements}
X.~T. Lee, A.~Khan, S.~Sen~Gupta, Y.~H. Ong, and X.~Liu, ``Measurements, analyses, and insights on the entire ethereum blockchain network,'' in \emph{Proceedings of The Web Conference 2020}, 2020, pp. 155--166.

\bibitem{akinshin2024trimmed}
A.~Akinshin, ``Trimmed harrell-davis quantile estimator based on the highest density interval of the given width,'' \emph{Communications in Statistics-Simulation and Computation}, vol.~53, no.~3, pp. 1565--1575, 2024.

\bibitem{jain1985p2}
R.~Jain and I.~Chlamtac, ``The p2 algorithm for dynamic calculation of quantiles and histograms without storing observations,'' \emph{Communications of the ACM}, vol.~28, no.~10, pp. 1076--1085, 1985.

\bibitem{usdtweth}
Etherscan, ``Uniswap v2 usdt-weth swapping pool,'' \url{https://etherscan.io/address/0x0d4a11d5eeaac28ec3f61d100daf4d40471f1852}, 2023.

\bibitem{breidenbach2021chainlink}
L.~Breidenbach, C.~Cachin, B.~Chan, A.~Coventry, S.~Ellis, A.~Juels, F.~Koushanfar, A.~Miller, B.~Magauran, D.~Moroz \emph{et~al.}, ``Chainlink 2.0: Next steps in the evolution of decentralized oracle networks,'' \emph{Chainlink Labs}, vol.~1, pp. 1--136, 2021.

\bibitem{abdk}
ABDK, ``Abdk libraries for solidity,'' \url{https://github.com/abdk-consulting/abdk-libraries-solidity}, 2019.

\bibitem{hardhat}
NomicFoundation, ``Hardhat,'' \url{https://github.com/NomicFoundation/hardhat}, 2024.

\bibitem{scikit-learn}
F.~Pedregosa, G.~Varoquaux, A.~Gramfort, V.~Michel, B.~Thirion, O.~Grisel, M.~Blondel, P.~Prettenhofer, R.~Weiss, V.~Dubourg, J.~Vanderplas, A.~Passos, D.~Cournapeau, M.~Brucher, M.~Perrot, and E.~Duchesnay, ``Scikit-learn: Machine learning in {P}ython,'' \emph{Journal of Machine Learning Research}, vol.~12, pp. 2825--2830, 2011.

\bibitem{xia2023detecting}
Q.~Xia, Z.~Huang, W.~Dou, Y.~Zhang, F.~Zhang, G.~Liang, and C.~Zuo, ``Detecting flash loan based attacks in ethereum,'' in \emph{2023 IEEE 43rd International Conference on Distributed Computing Systems (ICDCS)}.\hskip 1em plus 0.5em minus 0.4em\relax IEEE, 2023, pp. 154--165.

\bibitem{chen2024flashsyn}
Z.~Chen, S.~M. Beillahi, and F.~Long, ``Flashsyn: Flash loan attack synthesis via counter example driven approximation,'' in \emph{Proceedings of the IEEE/ACM 46th International Conference on Software Engineering}, 2024, pp. 1--13.

\bibitem{deng2024safeguarding}
X.~Deng, S.~M. Beillahi, C.~Minwalla, H.~Du, A.~Veneris, and F.~Long, ``Safeguarding defi smart contracts against oracle deviations,'' in \emph{Proceedings of the IEEE/ACM 46th International Conference on Software Engineering}, 2024, pp. 1--12.

\bibitem{qian2024mufuzz}
P.~Qian, H.~Wu, Z.~Du, T.~Vural, D.~Rong, Z.~Cao, L.~Zhang, Y.~Wang, J.~Chen, and Q.~He, ``Mufuzz: Sequence-aware mutation and seed mask guidance for blockchain smart contract fuzzing,'' in \emph{2024 IEEE 40th International Conference on Data Engineering (ICDE)}.\hskip 1em plus 0.5em minus 0.4em\relax IEEE, 2024, pp. 1972--1985.

\bibitem{zhong2024bitlink}
S.~Zhong and A.~Mueen, ``Bitlink: Temporal linkage of address clusters in bitcoin blockchain,'' in \emph{Proceedings of the 30th ACM SIGKDD Conference on Knowledge Discovery and Data Mining}, 2024, pp. 4583--4594.

\bibitem{elmougy2023demystifying}
Y.~Elmougy and L.~Liu, ``Demystifying fraudulent transactions and illicit nodes in the bitcoin network for financial forensics,'' in \emph{Proceedings of the 29th ACM SIGKDD Conference on Knowledge Discovery and Data Mining}, 2023, pp. 3979--3990.

\bibitem{bentley2022manipulating}
M.~Bentley, ``Manipulating uniswap v3 twap oracles,'' \url{https://github.com/euler-xyz/uni-v3-twap-manipulation/tree/master}, 2022.

\bibitem{park2023acon2}
S.~Park, O.~Bastani, and T.~Kim, ``Acon2: Adaptive conformal consensus for provable blockchain oracles,'' in \emph{Proceedings of the 32nd USENIX Conference on Security Symposium}, 2023, pp. 3313--3330.

\bibitem{huang2024advancing}
J.~Huang, L.~Kong, G.~Cheng, Q.~Xiang, G.~Chen, G.~Huang, and X.~Liu, ``Advancing web 3.0: Making smart contracts smarter on blockchain,'' in \emph{Proceedings of the ACM on Web Conference 2024}, 2024, pp. 1549--1560.

\bibitem{frassetto2023pose}
T.~Frassetto, P.~Jauernig, D.~Koisser, D.~Kretzler, B.~Schlosser, S.~Faust, and A.-R. Sadeghi, ``Pose: Practical off-chain smart contract execution,'' in \emph{Network and Distributed System Security (NDSS) Symposium 2023}, 2023, pp. 1--18.

\bibitem{kalodner2018arbitrum}
H.~Kalodner, S.~Goldfeder, X.~Chen, S.~M. Weinberg, and E.~W. Felten, ``Arbitrum: Scalable, private smart contracts,'' in \emph{27th USENIX Security Symposium (USENIX Security 18)}, 2018, pp. 1353--1370.

\bibitem{fang2022seframe}
M.~Fang, X.~Zhou, Z.~Zhang, C.~Jin, and A.~Zhou, ``Seframe: An sgx-enhanced smart contract execution framework for permissioned blockchain,'' in \emph{2022 IEEE 38th International Conference on Data Engineering (ICDE)}.\hskip 1em plus 0.5em minus 0.4em\relax IEEE, 2022, pp. 3166--3169.

\bibitem{li2019scalable}
C.~Li, B.~Palanisamy, and R.~Xu, ``Scalable and privacy-preserving design of on/off-chain smart contracts,'' in \emph{2019 IEEE 35th International Conference on Data Engineering Workshops (ICDEW)}.\hskip 1em plus 0.5em minus 0.4em\relax IEEE, 2019, pp. 7--12.

\bibitem{etherscanGas}
Etherscan, ``Ethereum gas tracker,'' \url{https://etherscan.io/gastracker}, 2024.

\bibitem{binance}
Binance, ``Bnb smart chain white paper,'' \url{https://github.com/bnb-chain/whitepaper}, 2020.

\bibitem{mclaughlin2023large}
R.~McLaughlin, C.~Kruegel, and G.~Vigna, ``A large scale study of the ethereum arbitrage ecosystem,'' in \emph{32nd USENIX Security Symposium (USENIX Security 23)}, 2023, pp. 3295--3312.

\bibitem{cao2021flashot}
Y.~Cao, C.~Zou, and X.~Cheng, ``Flashot: a snapshot of flash loan attack on defi ecosystem,'' \emph{arXiv preprint arXiv:2102.00626}, 2021.

\bibitem{angeris2022constant}
G.~Angeris, A.~Agrawal, A.~Evans, T.~Chitra, and S.~Boyd, ``Constant function market makers: Multi-asset trades via convex optimization,'' in \emph{Handbook on Blockchain}.\hskip 1em plus 0.5em minus 0.4em\relax Springer, 2022, pp. 415--444.

\bibitem{bordino2014stock}
I.~Bordino, N.~Kourtellis, N.~Laptev, and Y.~Billawala, ``Stock trade volume prediction with yahoo finance user browsing behavior,'' in \emph{2014 IEEE 30th International Conference on Data Engineering}.\hskip 1em plus 0.5em minus 0.4em\relax IEEE, 2014, pp. 1168--1173.

\end{thebibliography}

\end{document}